\documentclass[fleqn,usenatbib]{mnras}

\usepackage{newtxtext,newtxmath}
\usepackage[T1]{fontenc}

\DeclareRobustCommand{\VAN}[3]{#2}
\let\VANthebibliography\thebibliography
\def\thebibliography{\DeclareRobustCommand{\VAN}[3]{##3}\VANthebibliography}

\usepackage{graphicx}	% Including figure files
\usepackage{subcaption}

\title[Links Between Optical and X-ray Light in Scorpius X-1]{Links Between Optical and X-ray Light in Scorpius X-1}

\author[A. B. Igl et al.]{
Alexander B. Igl,$^{1}$\thanks{E-mail: aigl1@lsu.edu}
R. I. Hynes,$^{1}$
C. T. Britt,$^{2}$
K. S. O'Brien,$^{3}$
and V. J. Mikles$^{4}$
\\
$^{1}$Department of Physics and Astronomy, Louisiana State University, 202 Nicholson Hall, Tower Drive, Baton Rouge, Louisiana 70803, USA\\
$^{2}$Space Telescope Science Institute, 3700 San Martin Drive, Baltimore, Maryland 21218\\
$^{3}$Department of Physics, Durham University, Odgen Centre For Fundamental Physics West, Lower Mountjoy, South Rd, Durham DH1 3LE, United Kingdom\\
$^{4}$NASA Goddard Space Flight Center, Bldg L40, Greenbelt, MD 20770
}

\date{Accepted XXX. Received YYY; in original form ZZZ}

\pubyear{2021}

\begin{document}
\label{firstpage}
\pagerange{\pageref{firstpage}--\pageref{lastpage}}
\maketitle

\begin{abstract}
We observed the low-mass X-ray binary Sco X-1 for 12 nights simultaneously using the Rossi X-Ray Timing Explorer and the Otto Struve Telescope at McDonald Observatory at 1 second time resolution.  This is among the most comprehensive simultaneous X-Ray/optical data sets of Sco X-1.  Evidence of reprocessing was observed in the form of nine positive, near-zero lag peaks in the cross correlation function, eight of which were relatively small and took the shape of piecewise exponential functions.  These peaks were initially identified by eye, after which a computational identification scheme was developed to confirm their significance.  Based on their short lags (less than 4 seconds), as well as their occurrence on the flaring branch and soft apex, the small cross correlation features are likely to be caused by reprocessing off the outer disc, although the companion could still make a contribution to their tails.  The Z track was parameterized using a rank number scheme so that the system's location on the track could be numerically defined.  Plotting the results against the optical reveals an increasing step function when moving from the horizontal to the normal to the flaring branch, with differential optical levels at $\sim$0.47, $\sim$0.57, and $\sim$1.1 respectively.  An additional correlation between Z track location and the optical was found on the upper flaring branch.  An optical intensity histogram reveals a transition region between the normal and flaring branches with only intermediate fluxes.
\end{abstract}

\begin{keywords}
X-rays: binaries -- X-rays: individual: Scorpius X-1 -- accretion, accretion discs
\end{keywords}

%%%%%%%%%%%%%%%%%%%%%%%%%%%%%%%%%%%%%%%%%%%%%%%%%%

%%%%%%%%%%%%%%%%% BODY OF PAPER %%%%%%%%%%%%%%%%%%

\section{Introduction}

X-ray binary systems involve either a black hole or a neutron star accreting from a normal star or a white dwarf.  The material accretes through an accretion disc, emitting X-rays from the inner radii.  Depending on the mass of the companion, XRBs can be classified as either high mass ($>10M_ \odot$), low mass ($<1M_ \odot$), or intermediate mass.  Weakly magnetized low mass neutron star binary systems (NS-LMXBs) can be classified into two categories, based on their X-ray colour-colour diagram (CD) and hardness-intensity diagram (HID) behaviors: atoll and Z sources \citep{1989A&A...225...79H}.  Atolls generally have three separate components in their CDs, called the upper banana, lower banana, and island branches.  Z sources also have three components, called the horizontal, normal, and flaring branches (HB, NB, and FB respectively).  Atolls and Z sources tend to move through their tracks at different rates.  Atolls takes longer, tracing out the full path on the order of weeks, while Z sources can move through all three branches in hours or days \citep{1989A&A...225...79H}.

Differences between the two states extend beyond just the CDs.  Z sources are more luminous, often radiating near the Eddington limit, while atolls usually radiate at less than about 10\% of the Eddington limit.  Z sources tend to be much brighter in the radio band as well.  All Z sources have been detected in the radio, with Sco X-1 even having spatially resolved jets \citep{2001ApJ...558..283F}.  Both sources can experience type I X-ray bursts (thermonuclear burning of accreted material on the surface of the neutron star), but bursting behavior is much more frequent in atolls \citep{2008ApJS..179..360G}.  This is likely because the burning is more stable at near-Eddington luminosities in Z sources.

Z sources can be subdivided into two more groups, as first described by \citet{1994A&A...289..795K}.  The first of these are the Cyg-likes, named after Cyg X-2.  These tend to have a more traditional ``Z'' shape, and a more prominent HB.  Sco-likes (named after Sco X-1) are shaped like the Greek letter ``$\nu$'', with a shorter, upturned HB, and a long FB.  Several physical mechanisms have been suggested to try and explain the differences between Cyg-like, Sco-like, and atoll sources, including  inclination angles, NS spin rates, NS masses,  and NS magnetic field strengths \citep{1989A&A...225...79H}.  However, with the discovery of a single binary system \citep{2006ATel..696....1R}, most of these mechanisms have been disqualified.  XTE J1701-462 was the first LMXB to be observed transitioning between all three source types, moving from Cyg-like to Sco-like to atoll over the course of its 19 month outburst (\citeauthor{2007ATel.1144....1H}, \citeyear{2007ATel.1144....1H}; \citeauthor{2007ApJ...656..420H}, \citeyear{2007ApJ...656..420H}).  This suggests that the classes are driven only by accretion rate or flow geometry.

There are many theories as to what drives a system's position on its Z track.  It was long thought that the location on a CD was caused by a mass accretion rate that monotonically increases from HB to FB \citep{1995ApJ...454L.137P}.  More recent models have suggested other possibilities.  For example, \citet{2010ApJ...719.1350L} has an accretion rate that is unchanging throughout the Z track.  Another idea is given in \citet{2012A&A...546A..35C}, where the minimum accretion rate occurs at the soft apex and increases on either side.  Mass accretion rate is not directly observable, and its behavior with regards to the Z track is clearly model dependent. This raises the question of whether there is a measurable parameter that correlates in some way with the mass accretion rate.  Checking if the optical intensity satisfies this condition is a major goal of this research.

In addition to a binary system moving through these Z tracks, the tracks themselves can shift around the CD and change shape, a process known as secular drift.  Cyg-likes generally have the most obvious secular drift \citep{1996A&A...311..197K}, but it can be seen in Sco-likes as well.

Most of the optical flux in LMXBs comes from the reprocessing of X-rays by the outer disc or the companion star.  This leads to a delay in the optical signal (relative to the X-ray regime) on the order of the light travel time through the system ($\sim$10 s for LMXBs), based on the location of the reprocessing region.  Although the time for thermal reprocessing is not zero, it can usually be considered negligible relative to the light travel times \citep{1987ApJ...315..162C}.

The amount of light that is reprocessed can be described by the transfer function, which can be described by the equation,
\begin{equation}
    F_{\rm o}(t)=\int \Psi (\tau) F_{\rm x}(t- \tau)d \tau
\end{equation}
where $F_{\rm o}$ is the optical flux, $F_{\rm x}$ is the X-ray flux, $\Psi$ is the transfer function, $t$ is the time, and $\tau$ is the optical time delay.  \citet{2002MNRAS.334..426O} modelled the transfer function as a function of binary phase and time delay, using typical LMXB parameters.  Two major components came out of their model: The first was constant in phase, coming about from the outer disc contribution.  The second was quasi-sinusoidal, from the companion, and stretched to the highest reprocessing time delays possible within the system.  The companion contribution is much more dependent on parameters such as the binary separation and inclination angle, and so reprocessing off the companion can be used to further define these variables.  The quasi-sinusoidal component also contains the accretion stream, but the area for reprocessing is much smaller, and so the companion contribution dominates.  With a study like this one that utilizes both high time resolution data and energy-binned data, one could discriminate between the disc and companion reprocessing contributions using optical time delays, as well as identify the X-ray states during which reprocessing can be detected.

Sco X-1 is notable as the brightest persistent X-ray source in the sky (outside of the solar system), and the first discovered after the Sun \citep{1962PhRvL...9..439G}.  In addition, it is the closest Z source (permanent or transient), measured at a distance of 2.8 $\pm$ 0.3 kpc by \citet{1999ApJ...512L.121B} using Very Long Baseline Array parallax.  More recently, Gaia Early Data Release 3 parallax placed the system at 2.33$^{+0.12}_{-0.14}$ kpc \citep{refId0}.  Geometric parameters include an inclination of $\sim$$25^{\circ}$ \citep{2021MNRAS.508.1389C,2022ARep...66..348C} and a binary separation of $\sim$$3\times10^{11}$ cm, calculated from a mass ratio of 0.5, a NS mass of 1.4M$_\odot$ \citep{2015MNRAS.449L...1M}, and an orbital period of 18.9 hours \citep{1975ApJ...195L..33G,1975ApJ...201L..65C,2012ApJ...755...66H}.  V818 Sco is the optical counterpart to Sco X-1 \citep{1966ApJ...146..316S}, and its discovery allowed for X-ray/optical multiwavelength studies.  \citet{1967ApJ...150..851H} found that the Sco X-1 optical brightness had bimodal properties and could be split into clear high and low states.  \citet{1970A&A.....8....1H} later discovered trimodal properties in Sco X-1 as well.  \citet{1992A&A...265..177A} found similar bimodal behavior, and noted that the soft apex corresponded with the transition between the high and low states, which confirmed a similar suggestion by \citet{1986ApJ...306L..91P}.

A number of studies have found optical reprocessing in Sco X-1.  For example, \citet{2007MNRAS.379.1637M} found optical lags ranging from 5-10 seconds, and $\sim$11-16 seconds, which correspond to disc and companion reprocessing respectively (Bowen line filters were used to isolate the companion contributions).  Both instances of correlation occur in the FB (at the top for high lags and at the bottom for low lags).  The authors suggest that the large size of these lags could imply that either the NS mass or the inclination angle are bigger than previously reported.  \citet{2003MNRAS.345.1039M} also found lags in their optical/X-ray cross correlations that were bigger than expected based on the Sco X-1 geometry.  They propose that non-negligible thermal reprocessing times in the region of reprocessing could explain the lags.  With an extensive data set, one could answer the question of what conditions in the lightcurves are necessary for correlations, as well as find the prevalence of low lag (disc) and high lag (companion) peaks.  We present here an extensive data set spanning 12 nights to investigate these questions.

\section{Observations}
Simultaneous X-ray and optical observations of Sco X-1 were taken on the nights of May 19-30, 2009.  The observation periods are listed in Table \ref{table:tr}.  Much of the data overlapped in time, and so could be used to search for reprocessing on LMXB timescales.

The optical data were taken using the CCD photometer Argos on the Otto Struve Telescope at McDonald Observatory.  Only nine of the twelve nights allowed for observations, as thunderstorms prevented the dome from being opened on some nights.  The lightcurves were taken with 1 s time resolution, using a broad UBV filter (BG40) with a maximum transmittance at $\sim$500 nm wavelength and coverage of $\sim$350-650 nm.  Bias frames, dome flats, and dark frames were taken every night, and the images were processed using custom IDL programs written specifically for Argos.  Argos observed in 45 minute blocks to match RXTE visibilities.  Global Positioning System (GPS) 1 second ticks and Network Time Protocol (NTP) servers were used to sync the absolute time stamps.  All optical lightcurves used within cross-correlations were differential, which reduces atmospheric effects.  2MASS J16194990-1537248 was used as the comparison star for all of the data.
\begin{table}
    \centering
    \begin{tabular}{ll} 
        \hline
        Optical (UT) & X-Ray (UT) \\
        \hline
        May 19.73 - 19.91, 2009 & May 19.74 - 19.91, 2009 \\ 
        \hline
        May 20.78 - 20.89, 2009 & May 20.78 - 20.89, 2009 \\
        \hline
        May 21.70 - 21.70, 2009 & May 21.70 - 21.87, 2009 \\
        \hline
        N/A & May 22.68 - 22.90, 2009 \\
        \hline
        N/A & May 23.79 - 23.90, 2009 \\
        \hline
        May 24.71 - 24.87, 2009 & May 24.71 - 24.87, 2009 \\
        \hline
        May 25.75 - 25.80, 2009 & May 25.76 - 25.86, 2009 \\
        \hline
        N/A & May 26.75 - 26.84, 2009 \\
        \hline
        May 27.71 - 27.89, 2009 & May 27.72 - 27.89, 2009 \\
        \hline
        May 28.83 - 28.87, 2009 & May 28.83 - 28.87, 2009 \\
        \hline
        May 29.75 - 29.85, 2009 & May 29.75 - 29.85, 2009 \\
        \hline
        May 30.73 - 30.83, 2009 & May 30.73 - 30.83, 2009 \\
        \hline
    \end{tabular}
    \caption{Time ranges of all data.}
    \label{table:tr}
\end{table}

\begin{figure}
	\includegraphics[width=\columnwidth]{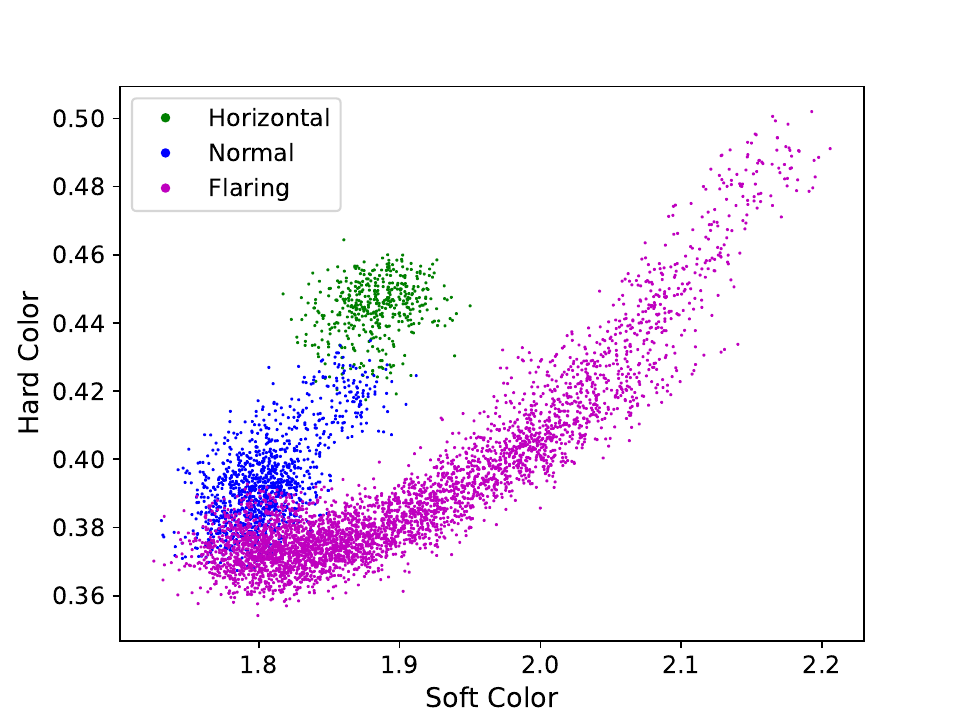}
    \caption{The Sco X-1 CD using STANDARD-2 data, which covers a full Sco-like Z track.}
    \label{fig:ccd}
\end{figure}

\begin{figure}
	\includegraphics[width=\columnwidth]{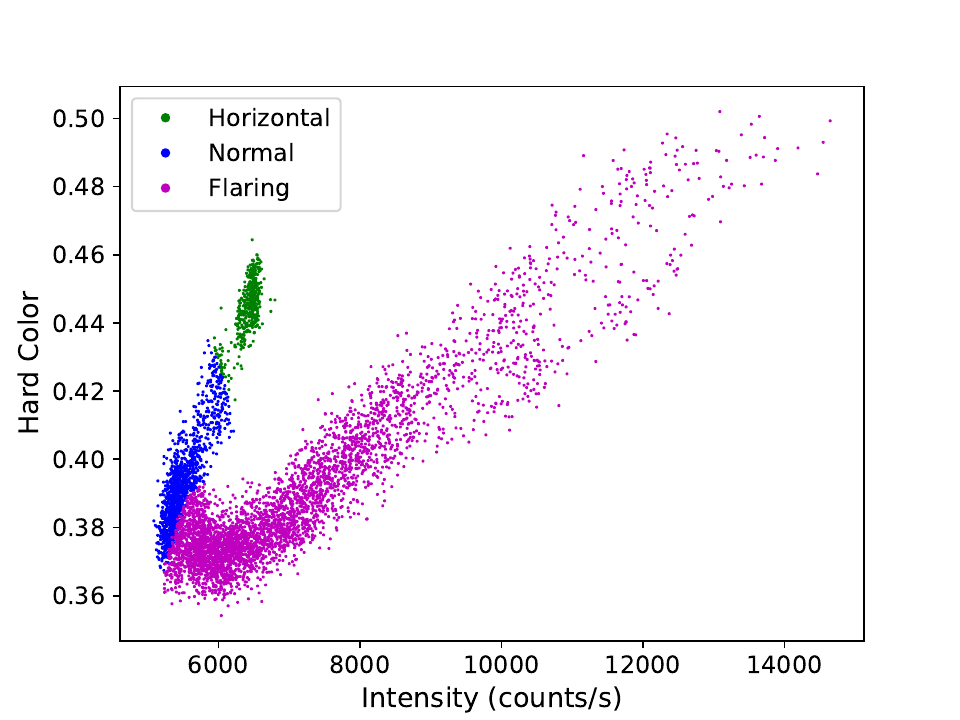}
    \caption{The Sco X-1 HID using STANDARD-2 data, which covers a full Sco-like Z track.}
    \label{fig:hid}
\end{figure}

The X-ray data were taken using the Rossi X-ray Timing Explorer (RXTE), using 100 kiloseconds of awarded time.  Because of the high X-ray brightness, images were taken with an offset ($\sim$22' RA, $\sim$37' DEC) to avoid instrument saturation. Light curves were extracted from the raw outputs using standard tools in HEASoft (version 6.29).  RXTE contained five proportional counter units (PCUs) on its proportional counter array (PCA).  The PCA had a total collecting area of 6500 cm$^2$, and a usable energy range of 2-60 keV.  As the instruments aged, certain PCUs began to discharge.  To prevent further damage, some of the PCUs were temporarily turned off for periods of time.  For this analysis, both the STANDARD-1 and STANDARD-2 data were taken only from PCU2, which allowed for a more consistent calibration.  STANDARD-1 data (no energy bands) were taken with a time resolution of 1 s, to match the resolution of the Argos data.  STANDARD-2 data were taken with a time resolution of 16 s, and energy bands of 2.0 - 4.0 keV, 4.0 - 6.5 keV, 6.5 - 9.0 keV, and 9.0 - 15.0 keV.  Soft colour is defined as the ratio of the fluxes in the 4.0 - 6.5 keV energy band over the fluxes in the 2.0 - 4.0 keV energy band.  Hard colour is defined as the ratio of the fluxes in the 9.0 - 15.0 keV energy band over the fluxes in the 6.5 - 9.0 keV energy band.  The data collected cover a full Sco-like Z track, as is clear in the corresponding colour-colour diagram (Fig. \ref{fig:ccd}) and hardness-intensity diagram (Fig. \ref{fig:hid}).

\section{Data Description}
\begin{figure}
	\includegraphics[width=\columnwidth]{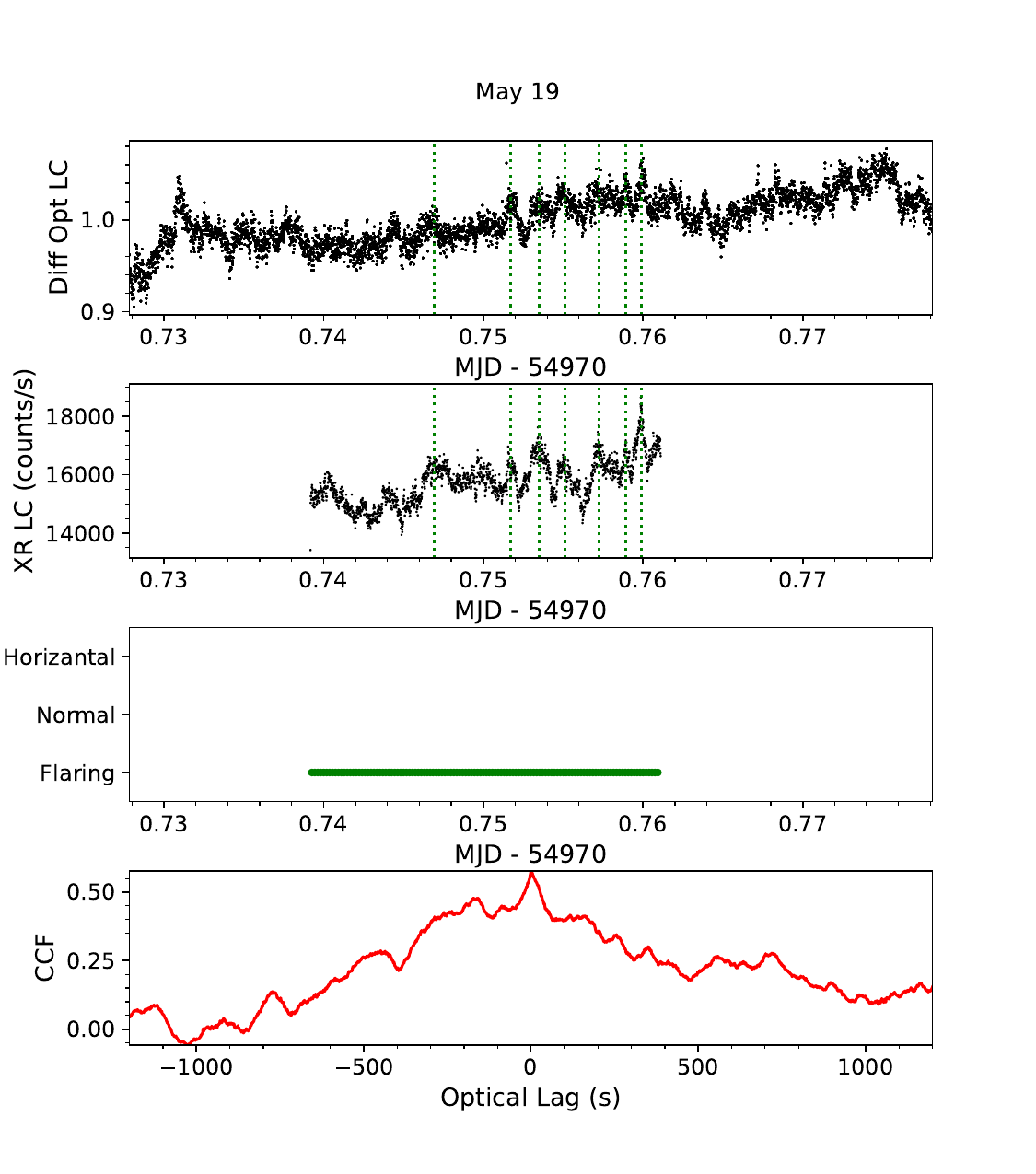}
    \caption{The most obvious of the potential reprocessing peaks.  Lines have been drawn in the lightcurves over flickers that are clearly being reprocessed.  The first plot is the differential optical lightcurve (normalized by the median), the second is the X-ray intensity, the third is the location on the Z track, and the last shows the unfiltered CCF.}
    \label{fig:britt_peak}
\end{figure}

The CD comprises a full Sco-like Z track (Fig. \ref{fig:ccd}).  Branch classification was performed using spline interpolations over the soft colour, hard colour, and intensity.  See Appendix \ref{sz_apx} for a detailed explanation of how these were calculated.  The HB is short and near-parallel along the hard colour axis, but clearly defined relative to the NB at lower hard colour values.  The FB is very prominent, extending over larger ranges of hard and soft colours than the HB and NB combined.  The FB also contains a ``kink'' near the midpoint, referenced in \citet{2005ApJ...623.1070M}.  Visually, it is the point where the FB begins to move to harder hard colours more quickly (i.e., the hard colour-soft colour slope increases).  While the HID also has the generalized Sco-like shape, the HB is much more difficult to discern from the NB than it was on the CD.  In addition, the kink is still present in the FB, but the higher intensity side has a noticeably higher spread than the lower intensity side.  The Z track undergoes no major secular drift over the 12 days Sco X-1 was observed.  As would be expected, the system generally moves through the NB before getting to the HB or FB, as opposed to skipping it entirely (Figs. \ref{fig:britt_peak} - \ref{fig:may27_blip} and Appendix \ref{nonblip_ccf_apx}).

The X-ray lightcurves are characterized by long timescale features combined with faster flickering (clear flickering can be seen in Fig. \ref{fig:britt_peak}, for example).  At some points in the lightcurves (FB and soft apex), there are bursts of intensity, combined with flickering that lasts between 1.5 and 7 minutes (see Figs. \ref{fig:may20a_blip} and \ref{fig:may27_blip} for clear examples).  This flickering can itself see the intensity rising up to $\sim$2000 counts.  This creates plateaus on top of the slow/flat features, which can themselves return back to the original intensity either gradually (Fig. \ref{fig:may27_blip}), or in an amount of time comparable to the initial burst (Fig. \ref{fig:may20a_blip}).  The heights of these intensity outbursts are seen to reach 50\% of the starting intensity, but are often smaller as well (10-20\%).  Similar flickering plateaus can be found in other observations as well, such as the lightcurves of \citet{2003MNRAS.345.1039M} and \citet{2008AIPC..984...15M}.  The large scale drops at the beginning and end of some X-ray observations (Fig. \ref{fig:may29c_full}) seem to be instrumental, as there is always a sharp intensity increase followed by a decrease at the beginning of a lightcurve, and a sharp intensity decrease at the end.  This is possibly due to RXTE slewing over the source to the offset position at the start, and then slewing away from the system altogether at the end.  It is likely that the standard good time intervals were insufficiently conservative in their selections.  The edges of these lightcurves were clipped before any further analysis was performed.

The optical lightcurves are normalized to the median of the intensities of Sco X-1, so that the differential flux values are not solely comparisons between Sco X-1 and a comparison star.  In general, they have a more varied range of behaviors than the X-ray lightcurves.  These include flickering of similar timescales to the X-ray flickering (Fig. \ref{fig:britt_peak}), constant intensities (Fig. \ref{fig:may27b_full}), slow rises/falls (Figs. \ref{fig:may19a_blip} and \ref{fig:may24c_blip}), sudden, short dips (Fig. \ref{fig:may24a_blip}), and small scale bursts on the order of 15 minutes (Fig. \ref{fig:may19b_blip}).  In many cases, these behaviors are unrelated to the X-ray lightcurve.  The X-ray flickering plateaus are often seen in the optical lightcurves as well.  They are nearly lined up with the X-ray plateaus, and sometimes are even more well defined when compared to other features.  This may imply that the optical has an X-ray reprocessed component.  There is one instance of an optical plateau unaccompanied by an X-ray counterpart (Fig. \ref{fig:may24a_blip}), but the flickering is much less pronounced.  Although the worst of the cloud effects were removed by using differential lightcurves and cutting out regions with the most extreme values, they can still be seen to a small degree in some regions (Fig. \ref{fig:may27b_full}).

\section{CCF Methodology}
\label{methodology}
Gaps in the optical will naturally occur due to issues such as cloudiness and bad data.  Because these create uneven sampling, the search for reprocessing was performed using discrete cross correlations (DCCFs), as in \citet{1988ApJ...333..646E}.  The DCCF takes the form,
\begin{equation}
    DCCF(\tau)=\frac{1}{M}\sum\limits_{i,j}\frac{(a_i-\bar{a})(b_j-\bar{b})}{\sigma_a\sigma_b}
\end{equation}
where $\tau$ is the lag, $a$ and $b$ are the lightcurves in question, $i$ and $j$ are the indices of data points within the lightcurves, $\sigma$ is the standard deviation of a lightcurve, and $M$ is the number of pairs to average over.  The DCCF offers an advantage over another commonly used version, the interpolated cross correlation \citep{1987ApJS...65....1G}, because it does not require the introduction of invented data when there are time gaps.

The X-ray data were broken into segments such that there were no data gaps larger than $\sim$10 minutes.  These data were then cross correlated with any optical observation they overlapped with in time.  The CCFs tend to be dominated by long timescale features, as seen by the figures in Appendix \ref{nonblip_ccf_apx}.  These structures can span on the order of 1000 seconds, making some of the shorter timescale features very difficult to detect by eye.  As such, high pass filters can be useful for finding smaller features.  For a discussion on how this is implemented, see Section \ref{blip_sect}.  Edge effects can be seen on some CCFs, where the magnitudes can reach extreme values at the maximum or minimum lightcurve overlap lag.  As none of these occur close to zero lag (the closest is at $\sim$500 s), they will have no bearing on analysis.

What qualifies as a CCF feature of interest depends on the process creating the feature and the geometric scales of the system.  For reprocessing, the optical signal will be received some time after the X-ray signal, which leads to a peak occurring at positive lags in the CCF.  The values of those lags are dependent on the light travel time through the binary system.  For LMXBs, reprocessing peaks will appear at up to tens of seconds of lag.  For Sco X-1, a potential event should be a peak at less than $\sim$17 s.

Out of all the CCFs, one peak in particular deserves focus (Fig. \ref{fig:britt_peak}).  It occurs at a lag of $\sim$1 s, and visually stands out when compared to the structure surrounding it, being taller than peaks of similar width.  Although no flickering plateaus appear in either lightcurve, there are some peaks in the optical lightcurve that correlate with X-ray peaks, such as the three peaks in the MJD region 54970.75-54970.76.  In addition, Sco X-1 is solidly on the FB during these observations, the only place reprocessing has been seen in Sco X-1 \citep{2016MNRAS.459.3596H}.  For these reasons, this peak is an extremely promising candidate for reprocessing.

\section{Minor Peaks within the CCFs}
\label{blip_sect}
Because the CCF peak described in Section \ref{methodology} showed up as a small peak superposed on a larger peak, we examined other CCFs more closely, and found other features, from here referred to as ``minor peaks'', which could be the result of reprocessing as well.  Minor peaks are defined as relatively small peaks at positive, near-zero lags.  They take the general shape of piecewise exponential functions, although the sharpness of the peaks is not consistent among them.  All of the minor peaks were initially identified by eye, via close examination of the CCFs.  Because these may or may not be much smaller than the scale of the surrounding CCFs, Butterworth filters \citep{Butterworth1930} were used to aid with visual identification.  Butterworth filters are designed to have a maximally flat response in the passband.  The response follows the formula
\begin{equation}
    H(j\omega)=\frac{1}{\sqrt{1+(\frac{\omega}{\omega_c})^{2n}}}
\end{equation}
where $j$ is the square root of -1, $\omega$ is the frequency, $\omega_c$ is the cutoff frequency divided by half of the sampling frequency, and $n$ is the filter order.  Filters for the visual identification of minor peaks were second order ($n$=2) and used cutoff times of 2 minutes, where cutoff time is the inverse cutoff frequency.  Filtering strips out features that span lags too long to signify reprocessing, resulting in easier identification.  These long-lag features tended to be comparatively large in magnitude, making the technique especially helpful for these data.  The filtered CCFs are what are used for the fits in Figs. \ref{fig:may19a_blip} - \ref{fig:may27_blip}, as well as the subsequent analysis.

\begin{figure*}
    \begin{subfigure}[b]{.475\textwidth}
        \setcounter{subfigure}{0}
        \includegraphics[width=\textwidth]{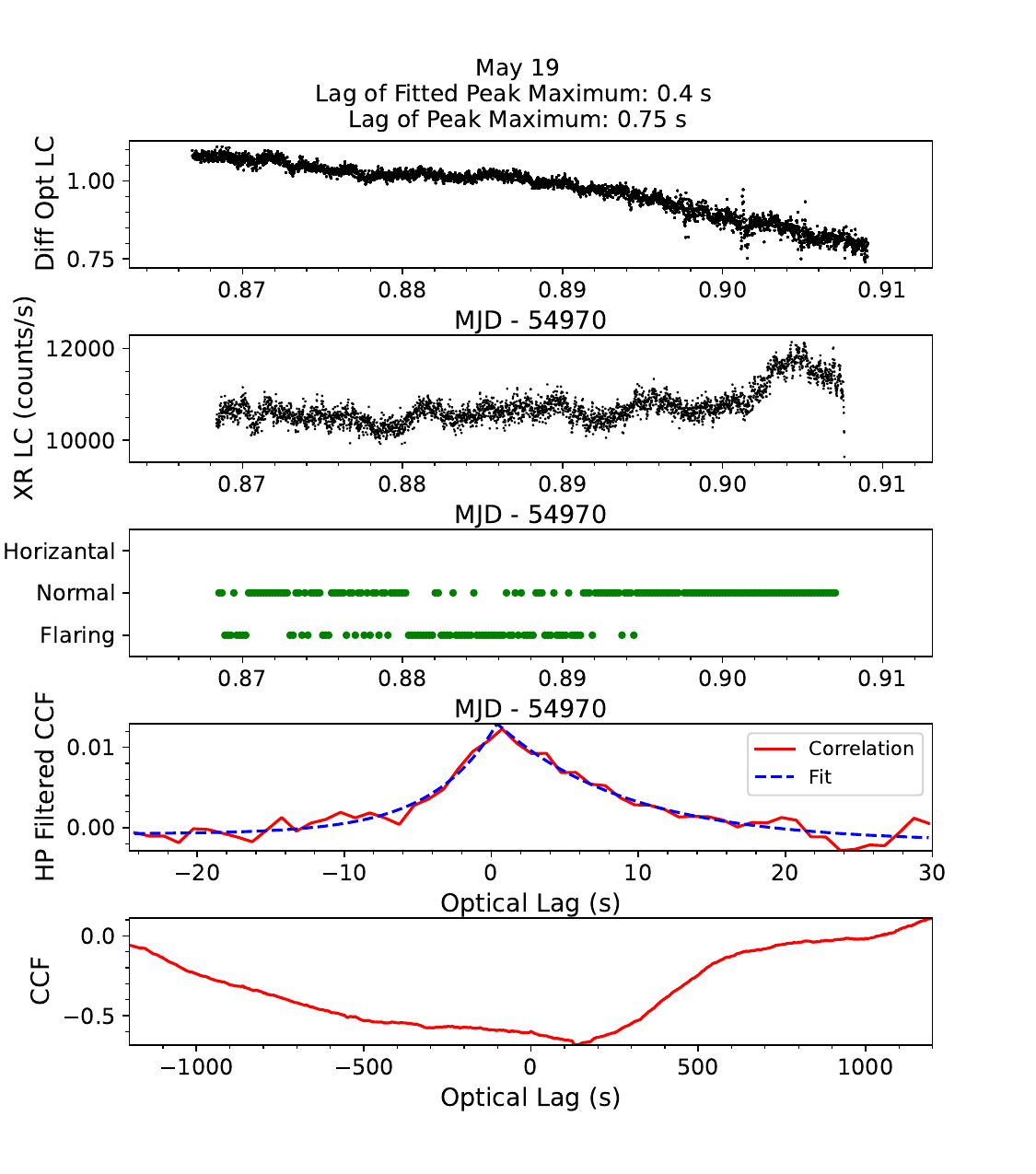}
        \caption[empty]{}
        \label{fig:may19a_blip}
    \end{subfigure}
    \hfill
    \begin{subfigure}[b]{.475\textwidth}
        \setcounter{subfigure}{2}
        \includegraphics[width=\textwidth]{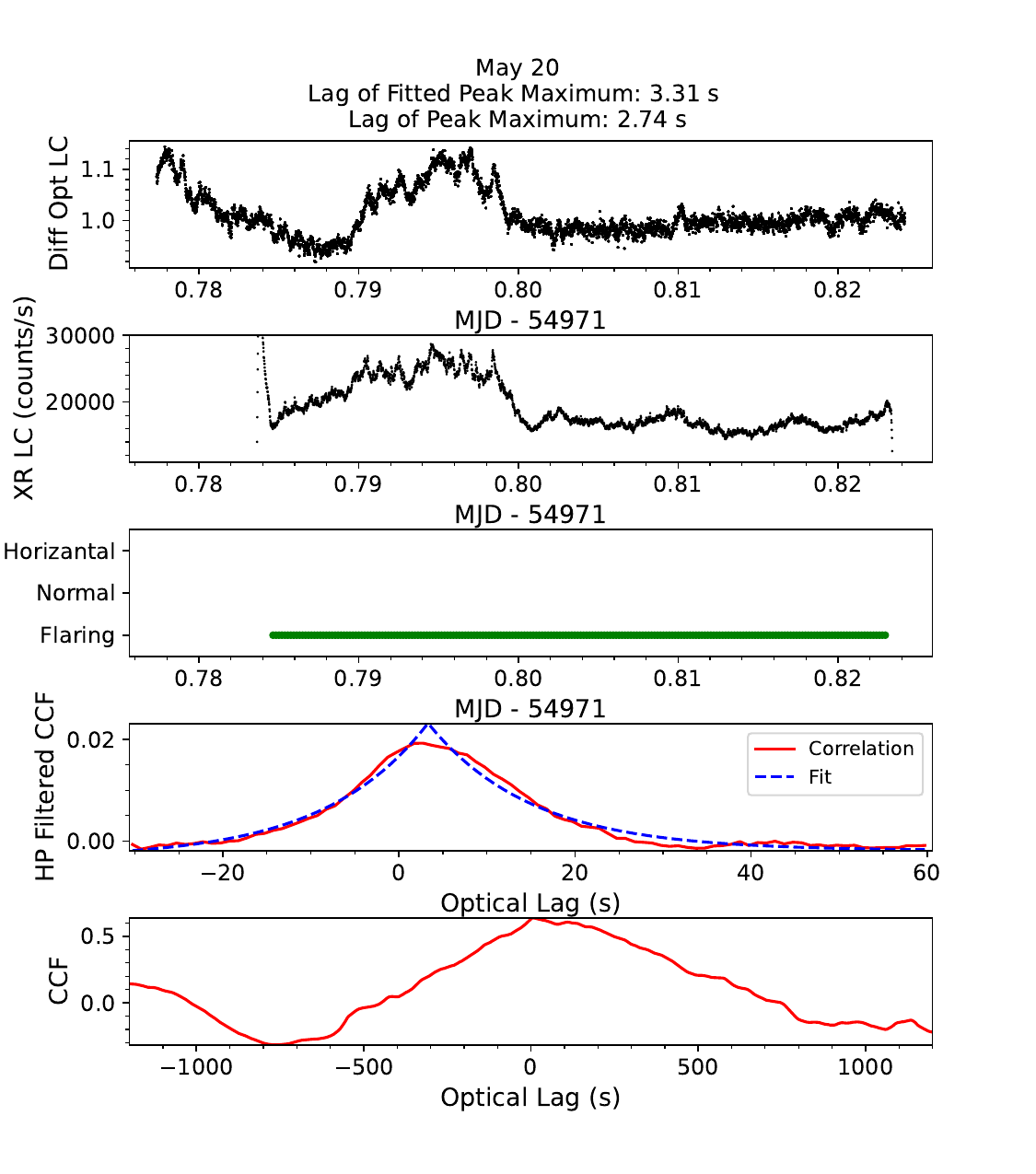}
        \caption[empty]{}
        \label{fig:may20a_blip}
    \end{subfigure}
    \vspace*{5mm}
    \begin{subfigure}[b]{.475\textwidth}
        \setcounter{subfigure}{1}
        \includegraphics[width=\textwidth]{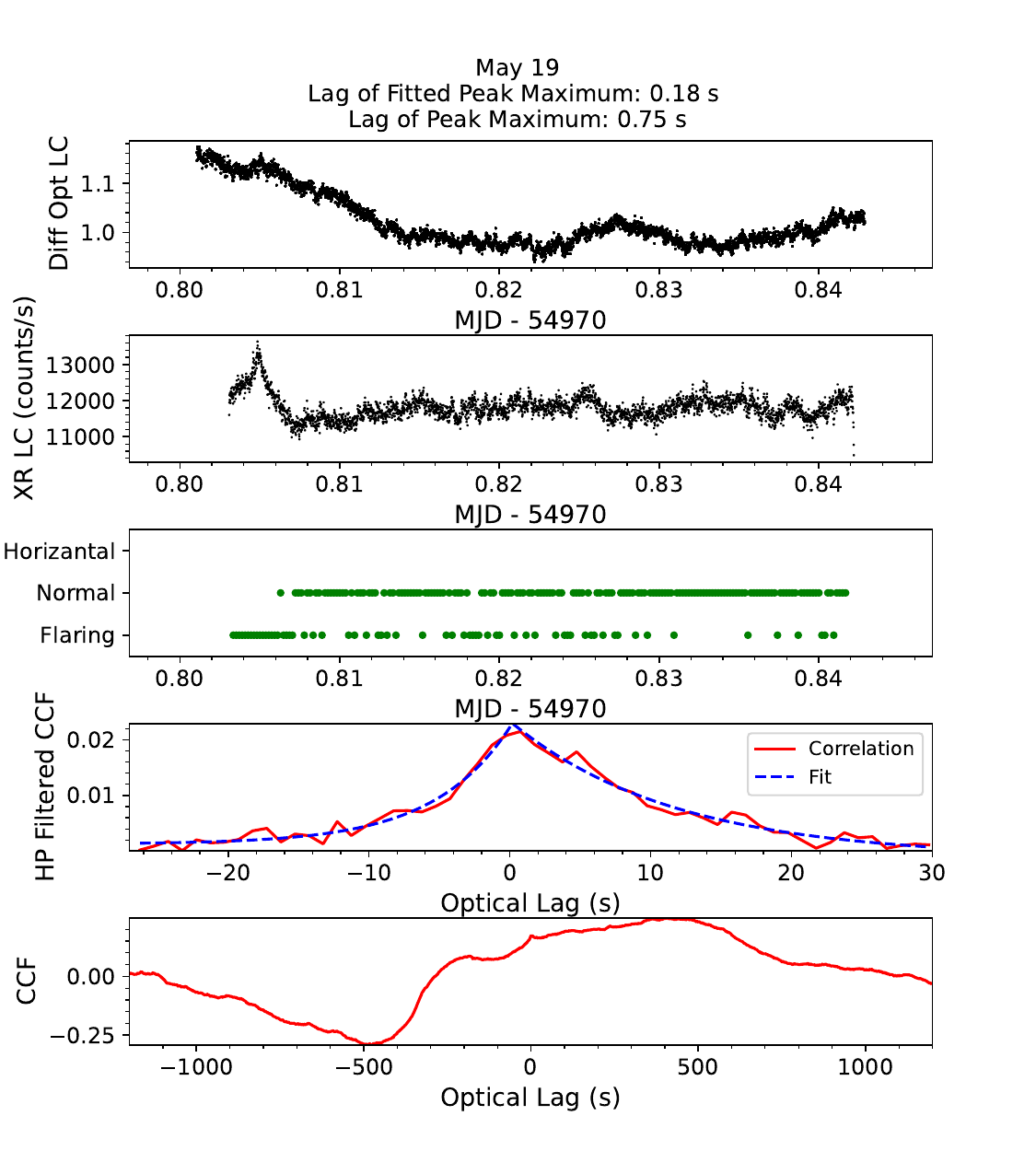}
        \caption[empty]{}
        \label{fig:may19b_blip}
    \end{subfigure}
    \hfill
    \begin{subfigure}[b]{.475\textwidth}
        \setcounter{subfigure}{3}
        \includegraphics[width=\textwidth]{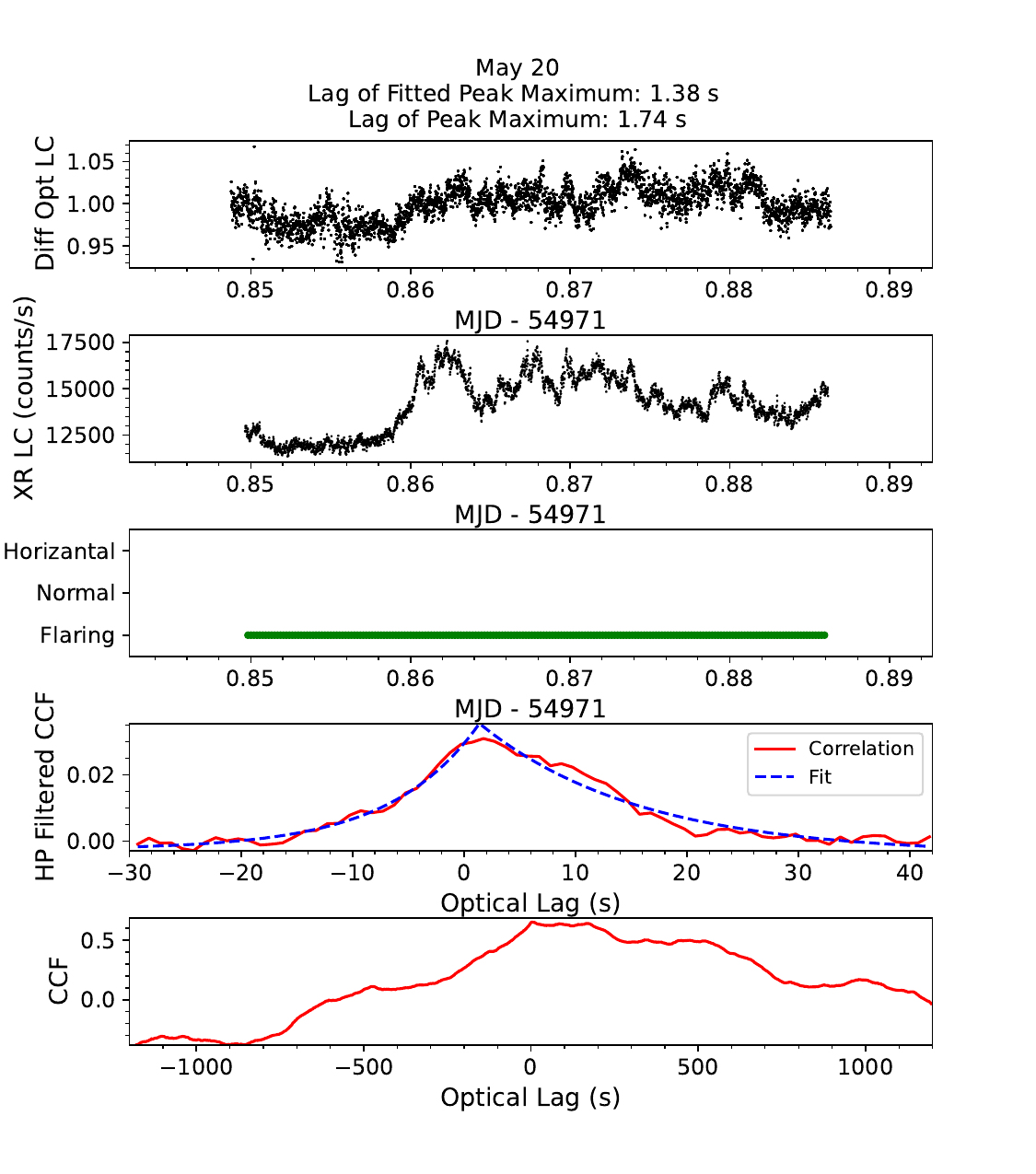}
        \caption[empty]{}
        \label{fig:may20b_blip}
    \end{subfigure}
    \caption{Examples of minor peaks and the data that produce them.  Note the similarities in optical and X-ray behavior, specifically the flickering plateaus in both.  The first plot is the differential optical lightcurve (normalized by the median), the second is the X-ray intensity, the third is the location on the Z track, the fourth shows the minor peak when the CCF has been run through a high pass Butterworth filter with a 2 minute cutoff time (red) and its fit to piecewise exponential functions (blue), and the last shows the unfiltered CCF.}
\end{figure*}

\begin{figure*}
    \ContinuedFloat
    \begin{subfigure}[b]{.475\textwidth}
        \setcounter{subfigure}{4}
        \includegraphics[width=\textwidth]{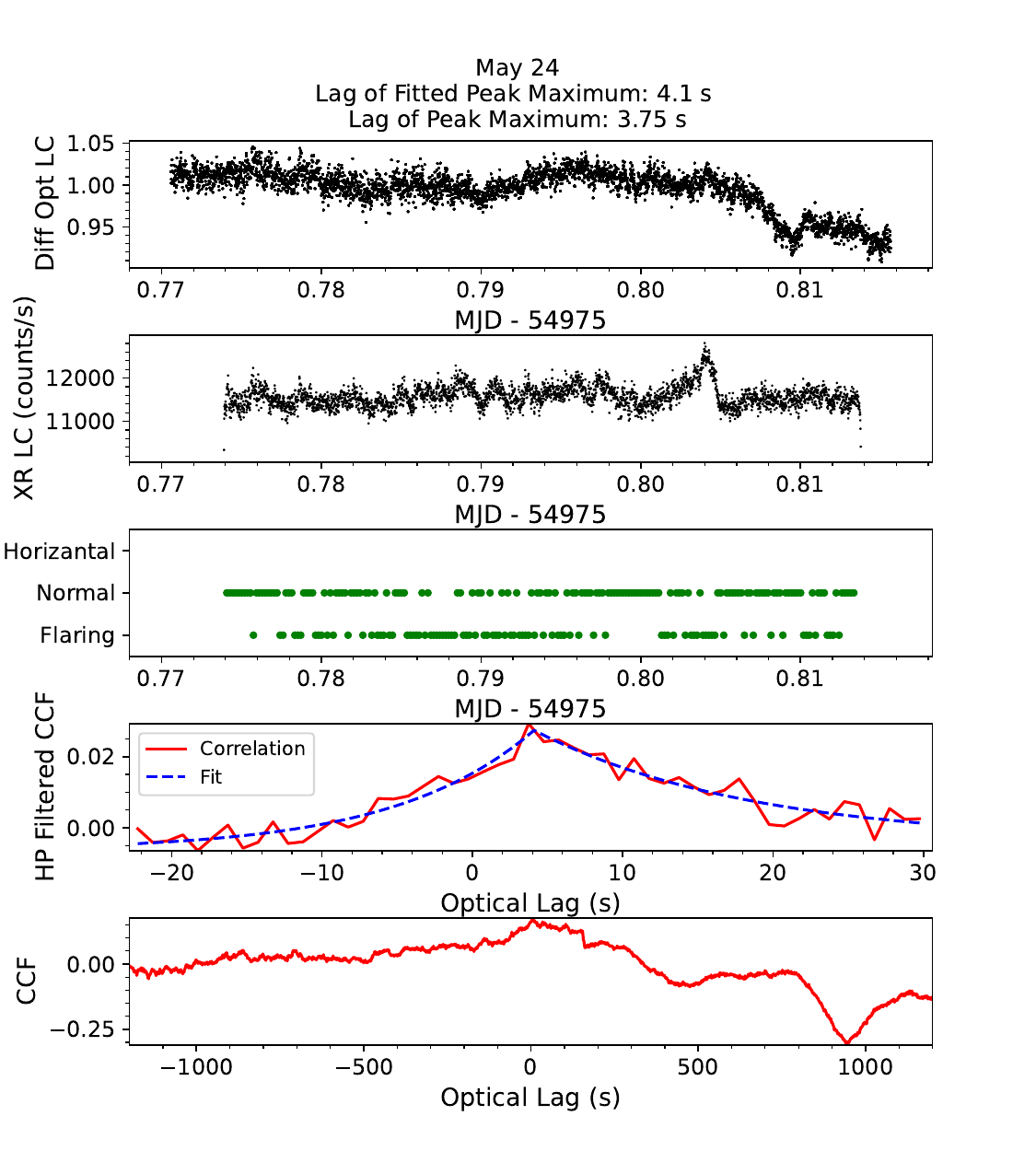}
        \caption[empty]{}
        \label{fig:may24a_blip}
    \end{subfigure}
    \hfill
    \begin{subfigure}[b]{.475\textwidth}
        \setcounter{subfigure}{6}
        \includegraphics[width=\textwidth]{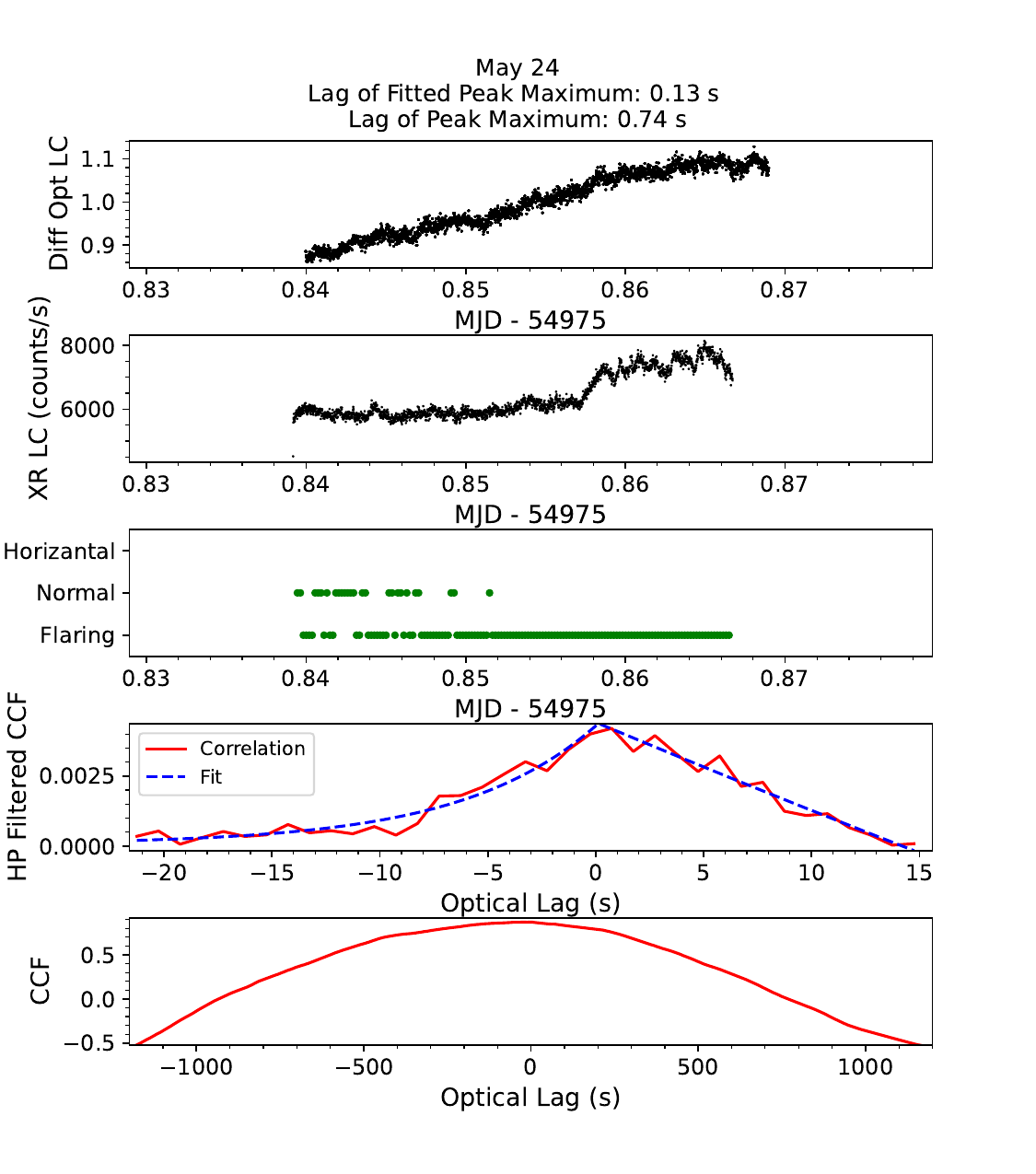}
        \caption[empty]{}
        \label{fig:may24c_blip}
    \end{subfigure}
    \vspace*{5mm}
    \begin{subfigure}[b]{.475\textwidth}
        \setcounter{subfigure}{5}
        \includegraphics[width=\textwidth]{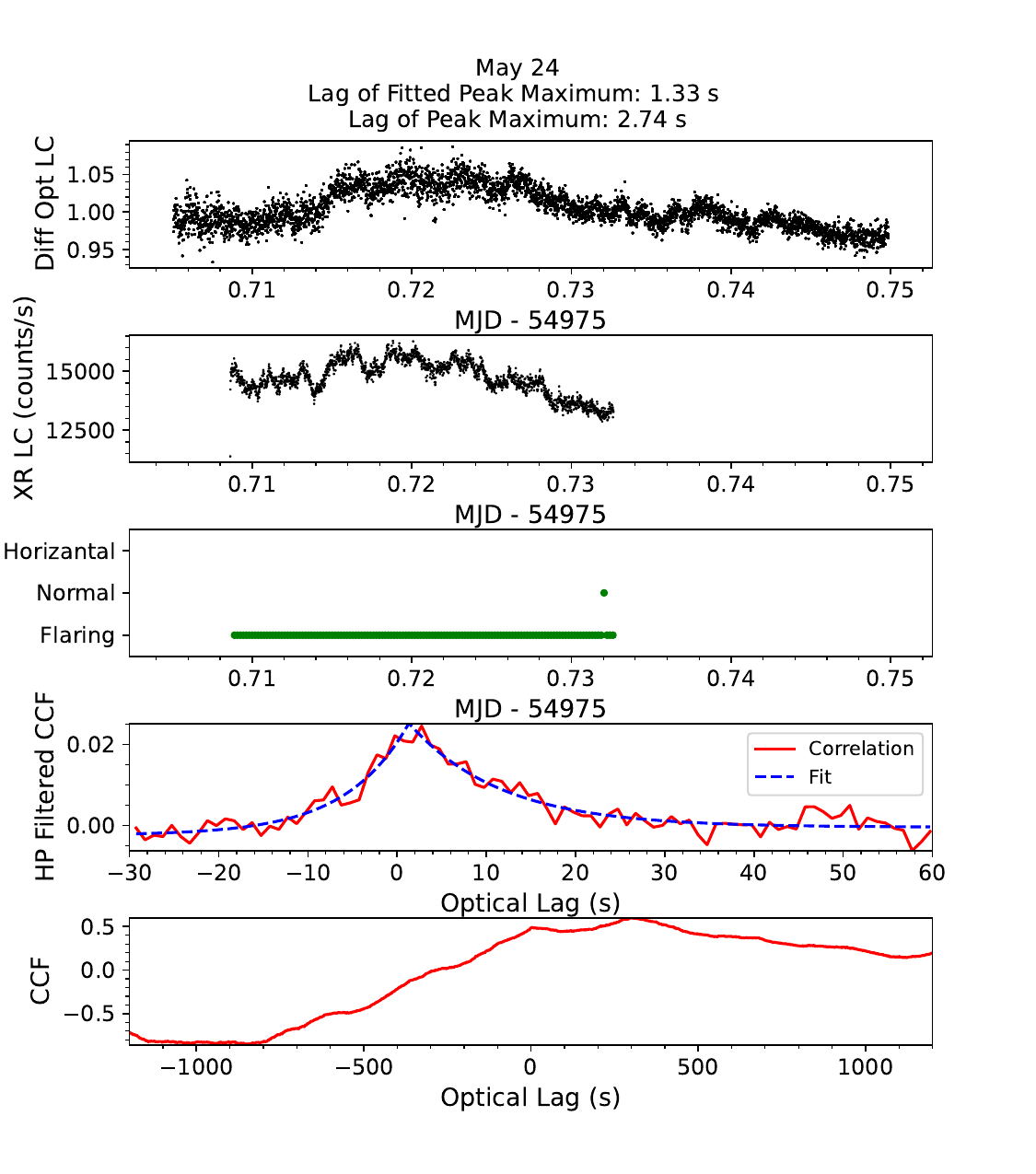}
        \caption[empty]{}
        \label{fig:may24b_blip}
    \end{subfigure}
    \hfill
    \begin{subfigure}[b]{.475\textwidth}
        \setcounter{subfigure}{7}
        \includegraphics[width=\textwidth]{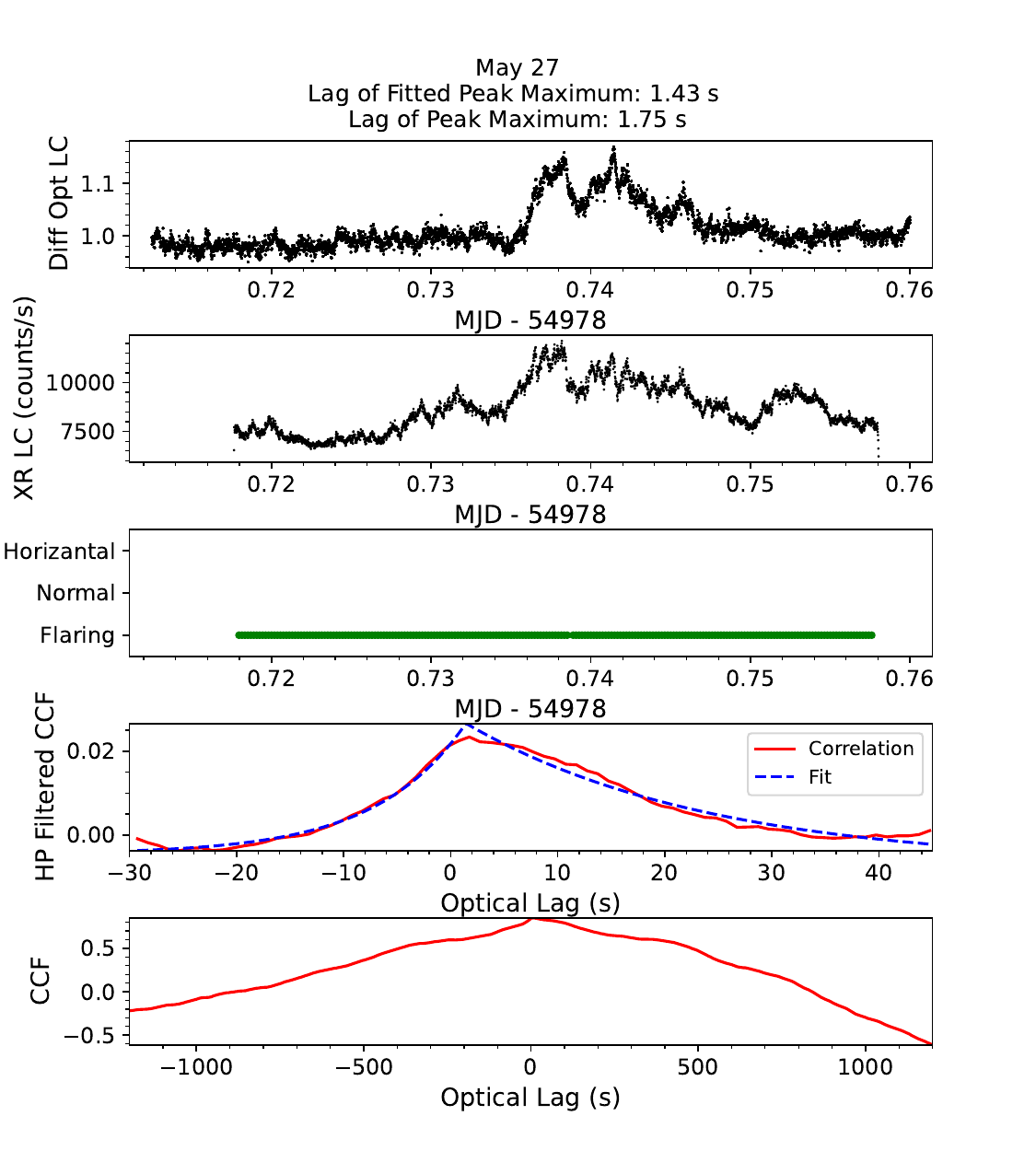}
        \caption[empty]{}
        \label{fig:may27_blip}
    \end{subfigure}
    \caption{(cont.) Examples of minor peaks and the data that produce them.  Note the similarities in optical and X-ray behavior, specifically the flickering plateaus in both.  The first plot is the differential optical lightcurve (normalized by the median), the second is the X-ray intensity, the third is the location on the Z track, the fourth shows the minor peak when the CCF has been run through a high pass Butterworth filter with a 2 minute cutoff time (red) and its fit to piecewise exponential functions (blue), and the last shows the unfiltered CCF.}
    \label{fig:blips}
\end{figure*}

\begin{figure}
	\includegraphics[width=\columnwidth]{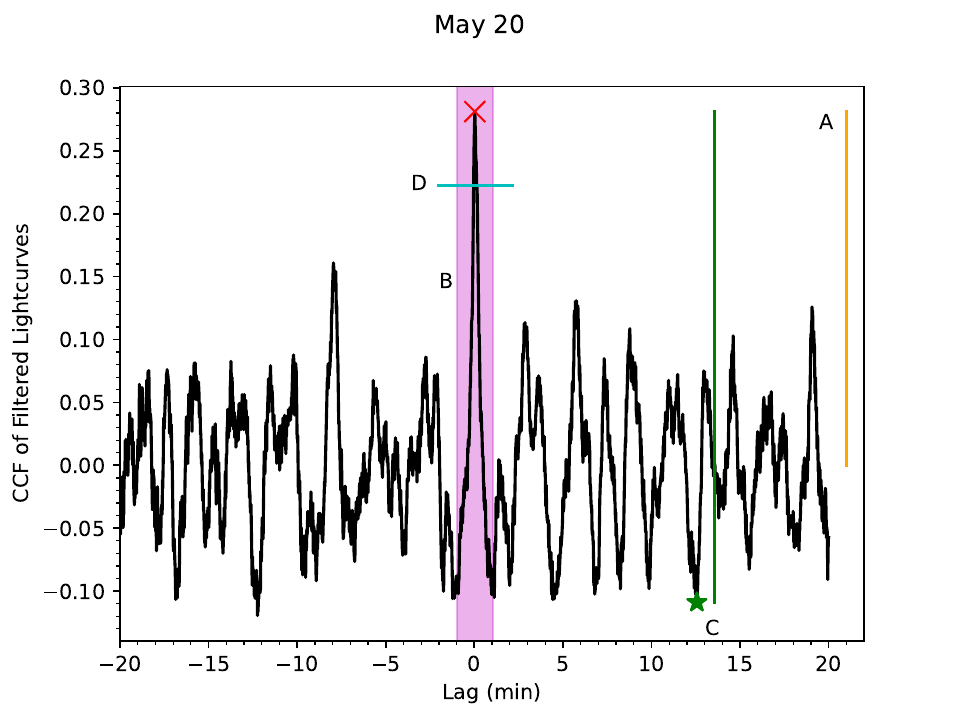}
    \caption{A demonstration of minor peak identification.  The minor peak maximum is represented by the red ``X'', the height is represented by the orange line (A), and the minimum distance between successive peaks is represented by the magenta shaded region (B).  The green star signifies the deepest trough between the peak and the edge of the optical lags.  The green line is the prominence, the difference between the peak and the trough.  This is calculated on both sides of the peak, with the true prominence being the smaller of the two (C).  The cyan line (D) represents the height where the width is calculated (15\% lower than the peak maximum).  The CCF is the result of Butterworth filtered lightcurves.}
    \label{fig:blip_id}
\end{figure}

A significant concern is to verify that the minor peaks are found at a statistically significant rate and are not coincidental.  For computational identification, we use a series of selection criteria that are summarized in Table \ref{table:blip_id}, in order of application (Fig. \ref{fig:blip_id}).  First, both the optical and X-ray lightcurves were high pass filtered using a Butterworth filter with a cutoff time of 3.6 min, to make the minor peaks more obvious.  Then, peaks were chosen based on: (a) the maximum being higher than a set value (peak height), (b) the distance between the maximum and the nearest maxima in lag, (c) the smallest difference between the peak maximum and the lowest contour bounded by the minor peak and the next higher peak or the edge of the data (prominence), and (d) the width of the peak (calculated at the \textit{relative height}).  This was done between $\pm$20 min of lag, so as not to include edge effects that were seen in the extreme lags of the CCFs.  It is important to note that these criteria were arbitrary peak properties that resulted in the identification of minor peaks already found by eye, and not necessarily based on prior expectations for the system.

To check the significance of the minor peaks, a rate of false occurrences should be calculated.  All optical and X-ray lightcurves that do not overlap in time were also cross correlated, leading to CCFs with lags that are much too large for reprocessing in LMXBs.  Because the identification scheme is independent of peak lag, all minor peaks found in these CCFs are therefore false occurrences.  On average, one false peak occurred every $\sim$158 min.  Poisson statistics, as well as the expectation that peaks would appear within $\pm$15 s, results in a vanishingly small probability that all ($\sim$10$^{-14}$), or even most ($\sim$10$^{-6}$ for half), of the minor peaks are false occurrences.  The chances of even one minor peak being a false positive is $\sim$7\%.  These may seem like optimistic numbers, as one might expect the rate of false occurrences to increase when looking only at FB data, due to the amount of activity and features in those sections of lightcurve.  However, the FB data only had a false peak every $\sim$136 min, not a meaningful change from the interval derived from the full data set.  The rate of false positives was a natural result of the selection criteria finding features similar to the visually identified reprocessing peaks exclusively.  Had the conditions been relaxed to find a wider range of peaks, the rate would have almost certainly increased.

\begin{table}
    \centering
    \begin{tabular}{lccr} 
        \hline
        XR High Pass Crit Time & 3.6 min \\
        \hline
        Opt High Pass Crit Time & 3.6 min \\ 
        \hline
        Height & 3.17 * S. Dev. \\
        \hline
        Dist. Between Peaks & 60 s \\
        \hline
        Prominence & Min: 0.3 | Max: 0.7 \\
        \hline
        Rel. Height & 0.15 \\
        \hline
        Width & Min: 3 s | Max: 20 s \\
        \hline
    \end{tabular}
    \caption{Conditions used for minor peak identification, in the order they were applied.  Note that for the relative height, 1 corresponds to the base of the peak at the lowest contour, and 0 represents the maximum of the peak.}
    \label{table:blip_id}
\end{table}

\begin{figure}
	\includegraphics[width=\columnwidth]{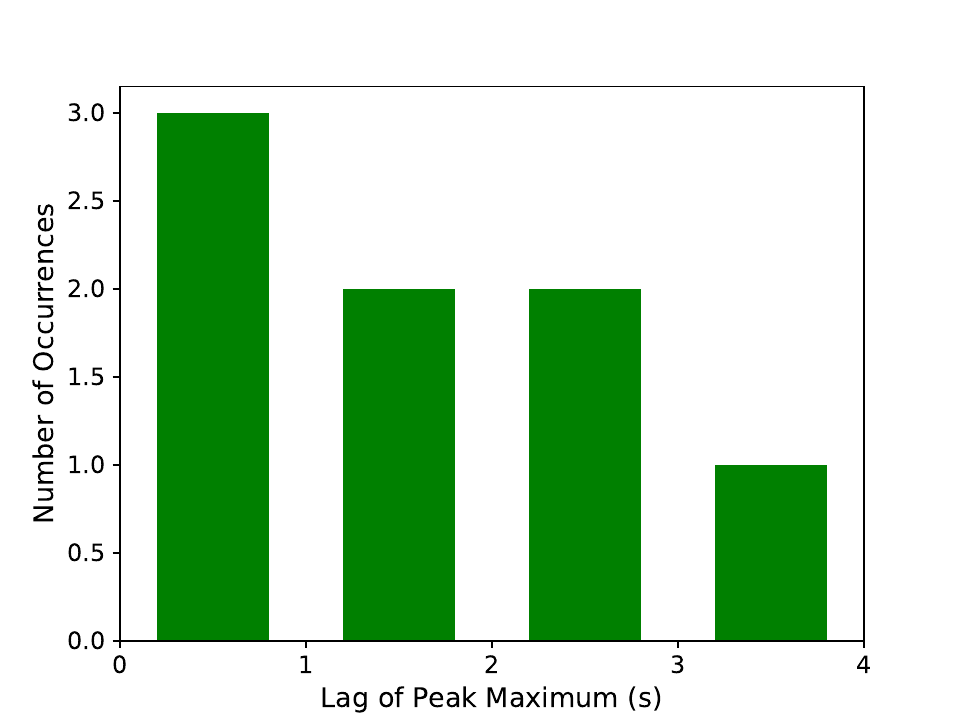}
    \caption{Histogram of peak lags for the eight smaller minor peaks.}
    \label{fig:lag_hist}
\end{figure}

A total of 8 minor peaks were identified (e.g., Figs. \ref{fig:may19a_blip} - \ref{fig:may27_blip}).  All peaked at lags of less than 4 s (Fig. \ref{fig:lag_hist}).  Initially, the peak referenced in Section \ref{methodology} was classified separately from the minor peaks, as it was much more prominent when compared to the surrounding CCF features.  However, when comparing the prominent peak and a stacked version of the minor peaks, they look very similar, implying that this peak is simply a stronger version of the minor peaks.  Although the minor peaks have unfiltered maxima at differing values (i.e., Figs. \ref{fig:may19b_blip} and \ref{fig:may20a_blip}), the filtered CCF minor peaks are comparable, as they identify correlations that are not necessarily the most obvious structure in the variability.  For all peaks (Fig. \ref{fig:britt_peak} included), the CCF behavior is more important than the actual magnitude.

\begin{figure}
	\includegraphics[width=\columnwidth]{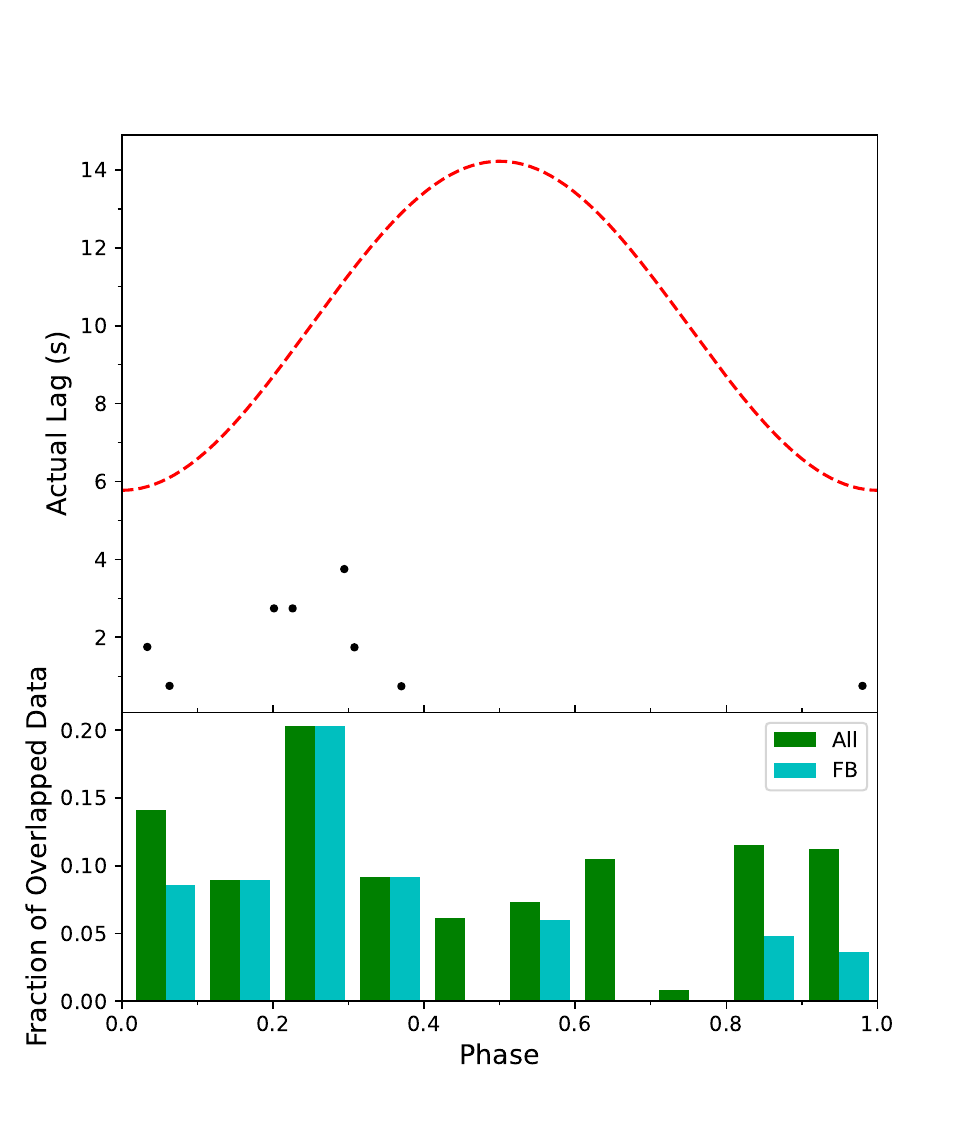}
    \caption{The top panel shows a phase-lag diagram of the eight detected minor peaks.  The red sinusoid is the expected lags for companion reprocessing based on the equation given in \citet{obrien_2000}.  The bottom panel is a histogram showing the distribution of data by phase as a fraction of the total amount of data.  In the low phase region containing the most minor peaks, Sco X-1 is almost always in the flaring branch ($s_z>2$).}
    \label{fig:lag_phase}
\end{figure}

Using the ephemeris from \citet{2014ApJ...781...14G}, the binary phase was calculated for each minor peak (Fig. \ref{fig:lag_phase}).  Note that a binary phase of 0 corresponds to the inferior conjunction of the system.  Lags due to companion reprocessing can be approximated by a sinusoid with mean $a/c$ and amplitude $(2a/c)sin(i)$ \citep{2002MNRAS.334..426O}, where $a$ is the binary separation, $c$ is the speed of light, and $i$ is the binary inclination. Due to the similarity of the shapes of minor peaks, it is likely that they are all reprocessing from a similar region of the system.  Based on the observed lags in Fig. \ref{fig:lag_phase}, this region is probably on the accretion disc.  The highest minor peak lag occurs at a phase of $\sim$0.3.  The sinusoid in Fig. \ref{fig:lag_phase} visualizes the expected companion lags (as calculated above), and shows that the minor peak would need to be $\sim$11 s to be companion reprocessing.  In fact, most of the points in Fig. \ref{fig:lag_phase} better coincide with reprocessing from the disc than the companion.

The top panel of Fig. \ref{fig:lag_phase} shows that most of the minor peaks occur at low phases, less than $\sim$0.4.  Although correlations from companion reprocessing would become more likely as the irradiated face comes into view, one would not expect these correlations to be so asymmetric about the orbit.  The bottom panel of the same figure shows the distribution of data by phase as a percentage of the total.  Most of the data with phases of less than 0.5 are in the FB ($\sim$80\%), while less than half is in the FB for phases greater than 0.5.  In addition, there is more FB data when the phase is less than 0.5 ($\sim$75\% of the total).  Looking at these differences in coverage, it is not altogether unexpected that this distribution of data would lead to most of the minor peaks occurring at the smaller phases.  

\begin{figure}
	\includegraphics[width=\columnwidth]{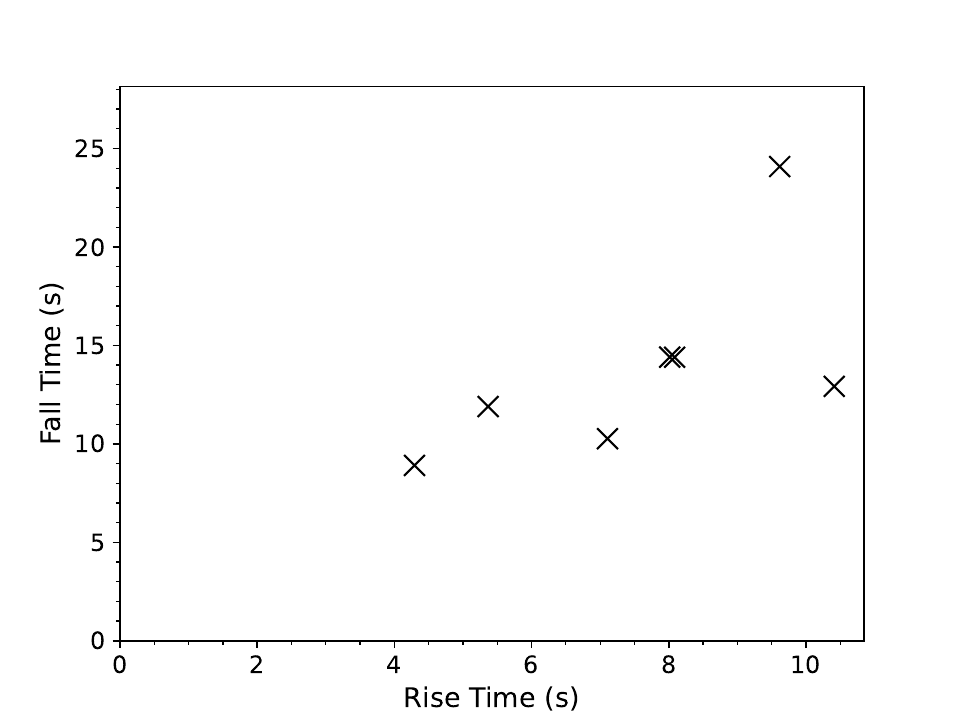}
    \caption{A comparison of minor peak rising lag scales and falling lag scales, based on piecewise exponential fits.  One minor peak with a falling lag scale of $\sim$200 s is not included in the plot.}
    \label{fig:rise_fall}
\end{figure}

Analysis of the CCF exponential fits reveals that the rising timescales of the minor peaks tend to be smaller than the falling timescales (Fig. \ref{fig:rise_fall}).  Here, the rising and falling timescales are the inverse rate of growth from the exponential fits.  Using a two sample Kolmogorov-Smirnov test (excluding one data point with a fall time of $\sim$200 s) yields a P-value of 0.05, making it unlikely that the samples could have been drawn from the same distribution.  In addition, Fig. \ref{fig:rise_fall} shows the rising minor peak times plotted against the falling minor peak times (where ``times'' are the inverse exponents), disregarding an outlier with a fall time of almost 200 s.  The gradient of the fit line is greater than 1, illustrating that the minor peaks are asymmetric.  The Pearson correlation coefficient of these times are 0.63, with a p-value of 0.13.  The fast rise/slow decay shape of the minor peaks are a natural result of transfer function properties.  Also of note is the magnitude of these times, ranging from $\sim$5-10 s.  These are similar to the lag times one would see from companion reprocessing, and so companion reprocessing cannot be completely ruled out; it may contribute to the smearing of the response, even if not to the peak.  A comparison of autocorrelation central peak widths and the rising/falling times of the minor peaks is dominated by scatter and does not convincingly show that the steepness is driven by the X-ray lightcurve timescales.

An additional effort was taken to search for trends involving the first four moments (mean, variance, skew, kurtosis) of the minor peaks.  Various quantities were plotted against the moments, including hard colour, soft colour, flux, phase, position on the CD Z track ($s_z$), and the rate of change of Z track position ($ds_z/dt$).  However, none of the quantities revealed any sort of tendencies or correlations that were not already obvious.  They do, on the other hand, further illustrate the asymmetry in the minor peak skews (moment 3) seen in Fig. \ref{fig:rise_fall}, as all of the skews are positive, corresponding to a faster rise than fall.  In addition, the asymmetry reinforces that the minor peaks are real, and not statistical coincidences, as random peaks would not have systematic lopsidedness.  This behavior has precedent, as in the CCFs of \citet{2009MNRAS.399..281H}, in their echo mapping of the black hole XRB Swift J1753.5–0127.  These correlations were attributed to thermal reprocessing on the disc, and behaved similarly to CCFs resulting from convolving model transfer functions and X-ray ACFs \citep{2002MNRAS.334..426O,2003MNRAS.345..292H}.

\section{Spectral Behaviors}
\begin{figure}
    \includegraphics[width=\columnwidth]{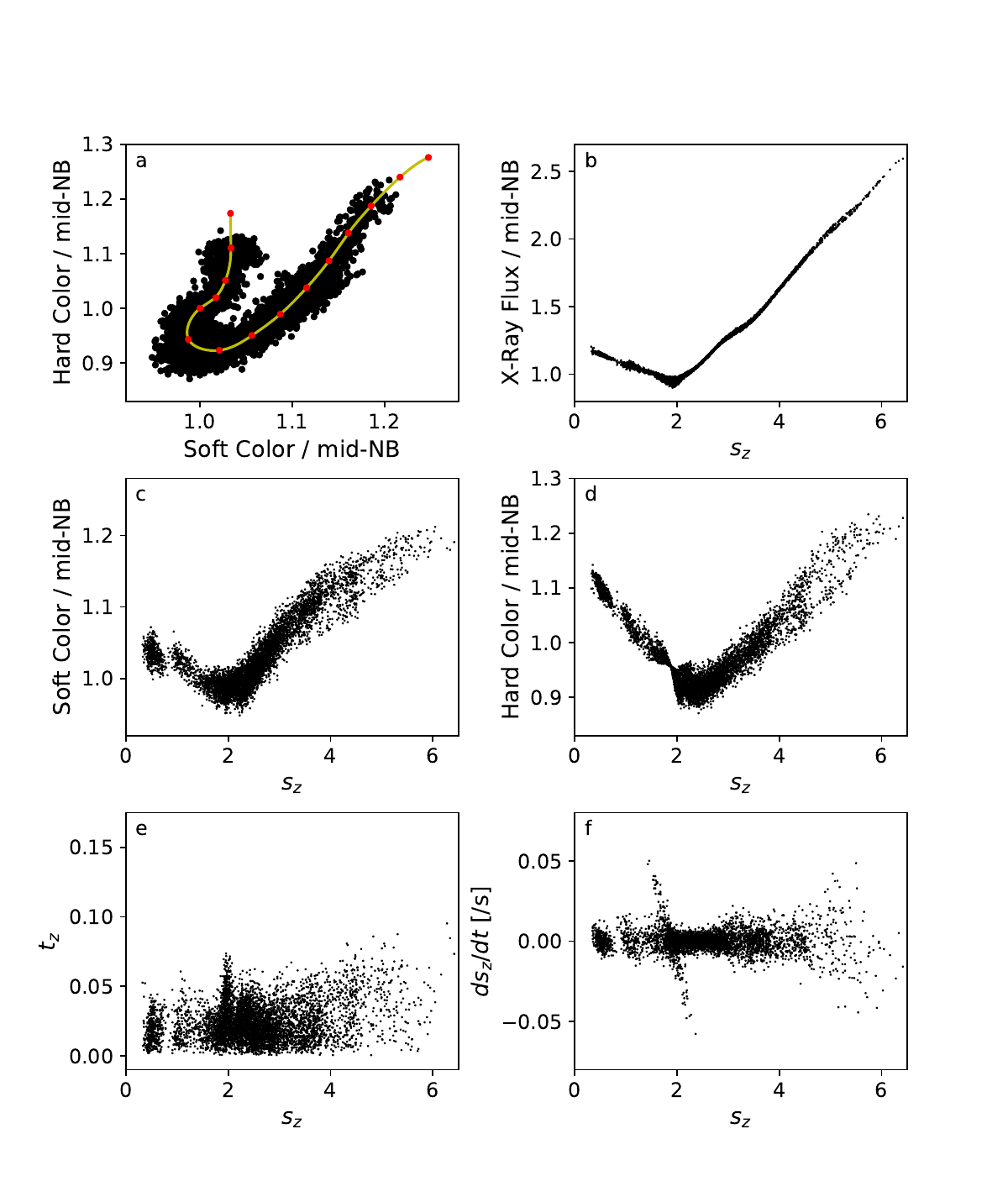}
    \caption{The results of the 3D spline fit, resulting in ($s_z$,$t_z$) coordinates for the Sco X-1 data}
    \label{fig:scox1_st}
\end{figure}

\begin{figure}
    \includegraphics[width=\columnwidth]{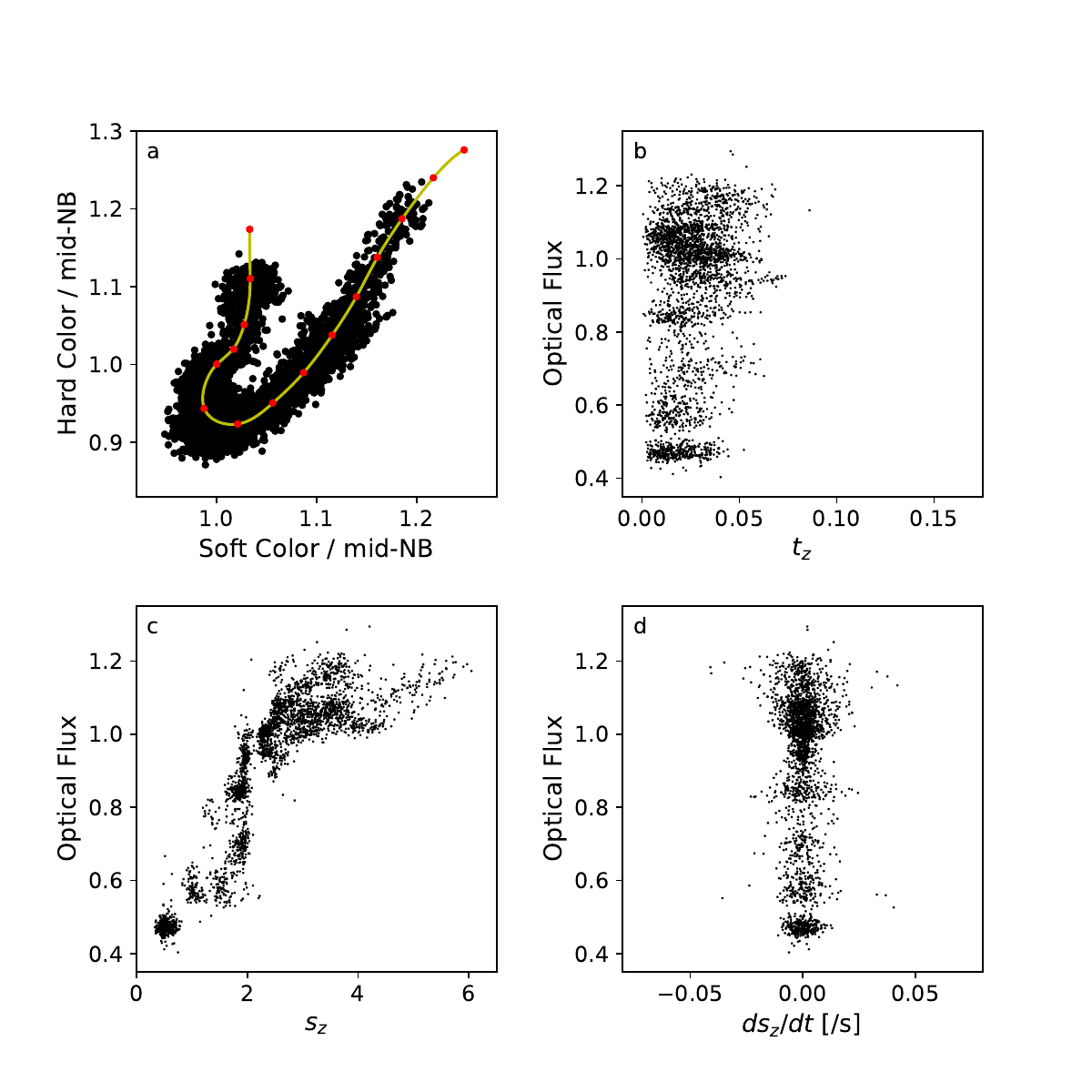}
    \caption{The ($s_z$,$t_z$) variables plotted against the Sco X-1 differential optical data (divided by the median).}
    \label{fig:scox1_st_optical}
\end{figure}

\begin{figure}
    \includegraphics[width=\columnwidth]{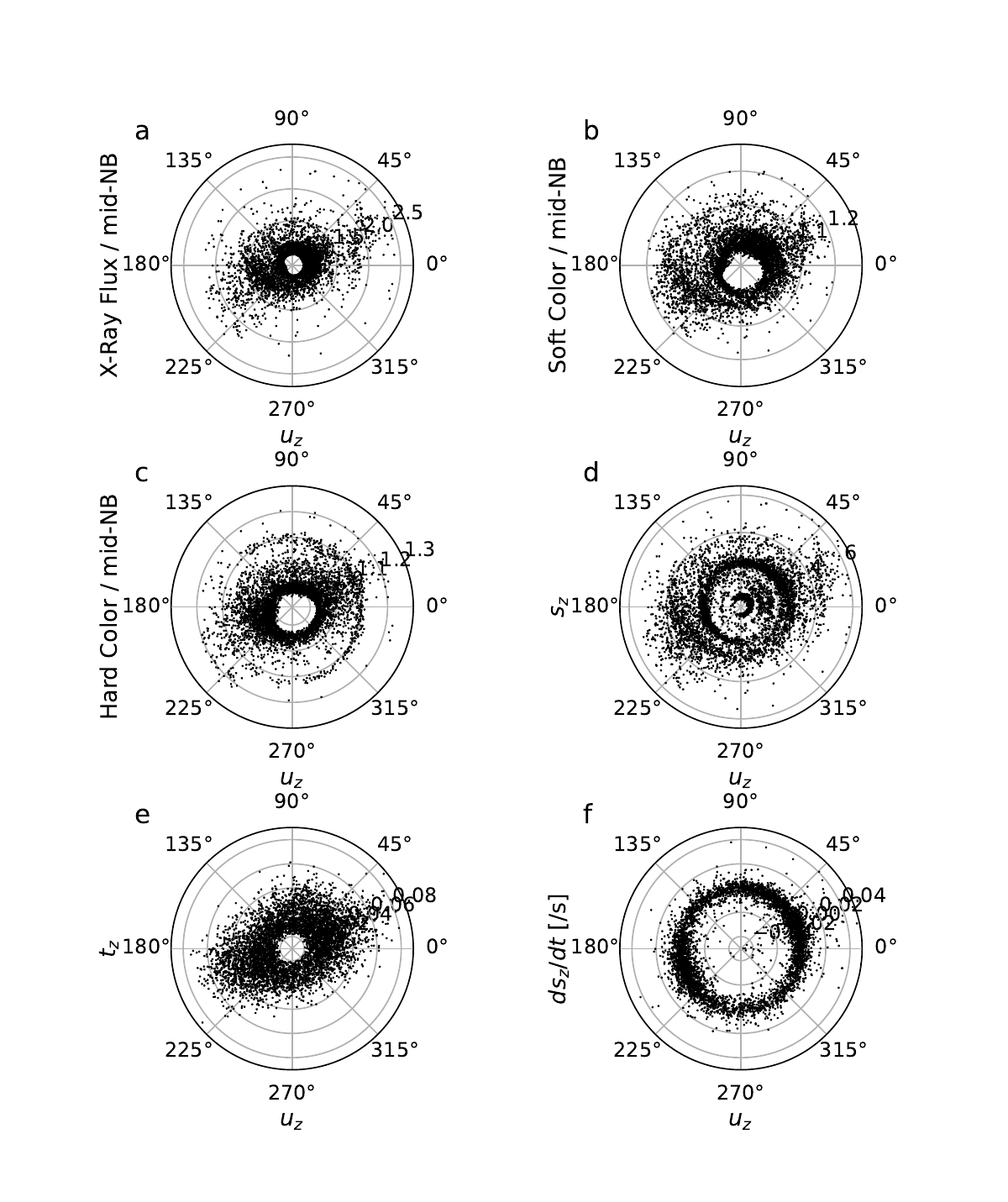}
    \caption{The $u_z$ variable plotted against a variety of others.}
    \label{fig:scox1_uz}
\end{figure}

 We mapped the data points into a coordinate system relative to the Z track defined by the parameters $s_z$, $t_z$, and $u_z$ (see Appendix \ref{sz_apx} for in depth discussion of how the mapping was performed).  $s_z$ is defined as the location along the Z track, $t_z$ is the distance away from the Z track, and $u_z$ is the angular component around the Z track.  The coordinates ($s_z$, $t_z$, $u_z$) can then be thought of as a cylindrical coordinate system where the height axis does not have a static direction.  $s_z=1$ and $s_z=2$ are defined as the hard and soft apexes respectively, and the scaling of $s_z$ in the HB and FB is based on the scale of the NB (between $s_z=1$ and $s_z=2$).  This means that $s_z$ does not have set minimum or maximum values.  The results of the ($s_z$, $t_z$, $u_z$) mapping can be seen in Figs. \ref{fig:scox1_st}, \ref{fig:scox1_st_optical}, and \ref{fig:scox1_uz}.  For the mapping of these data, the beginning of the HB is near $s_z=0.5$, and the end of the FB approaches $s_z=6$, making the FB midpoint about $s_z=4$.

\begin{figure}
	\includegraphics[width=\columnwidth]{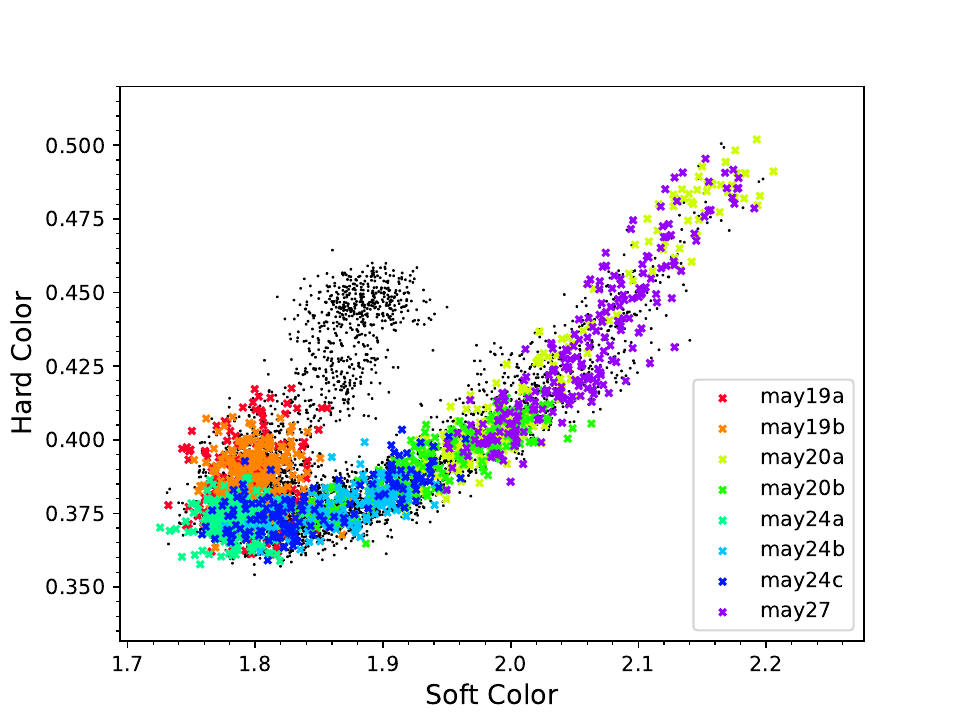}
    \caption{The Sco X-1 CD with minor peak locations marked in colour (see Section \ref{blip_sect} for the definition of a minor peak).  While it appears as though some of the minor peaks stretch well into the normal branch (specifically on May 19), this is only due to some outlier points.  The densest clusters of those points lie on the soft apex.}
    \label{fig:ccd_blips}
\end{figure}

Plots b, c, and d in Fig. \ref{fig:scox1_st} show $s_z$ plotted against the observed variables used for the ($s_z$,$t_z$,$u_z$) mapping, with a good correlation with respect to each.  As one might expect for a Sco-like with a near-vertical HB, all of the variables increase from the soft apex, except for the soft colour in the HB.  $s_z$ is also plotted here against $t_z$ (Fig. \ref{fig:scox1_st}, plot e), with no correlation apparent, although there is a noticeable increase in scatter in the upper half of the FB.  Plot f shows $ds_z/dt$ plotted against $s_z$ itself, and also has a larger spread in the upper FB.  The effects of this behavior can be seen in Fig. \ref{fig:ccd_blips}.  Even though the time ranges of the lightcurves used to create the minor peaks are similar lengths, the minor peak locations on the upper FB cover much more of the Z track than the minor peaks on the lower FB or the soft apex.  The diagonal line of points centered around $s_z=2$ in plot f is not a real trend, but rather an artifact from the mapping of points on the concave side of the Z diagram.

Plot c in Fig. \ref{fig:scox1_st_optical} shows four clear regions in the $s_z$ vs. optical plot: an uncorrelated step in the HB at $\sim$0.47, an uncorrelated step in the NB at $\sim$0.58, an uncorrelated step in the lower FB at $\sim$1.1 (with a much higher dispersion than the HB and NB steps), and a positive correlation in the upper FB ($s_z>4$) with a slope of 0.077.  Note that this final region is also associated with the increased spread in $t_z$ and $ds_z/dt$ described above.  The soft apex ($s_z=2$) shows a large range of fluxes among the NB/FB steps, but no such spread exists at the hard apex ($s_z=1$).  This is possibly more to do with the amount of data obtained around the hard apex, which is significantly less than exists on the soft.

\begin{figure}
    \includegraphics[width=\columnwidth]{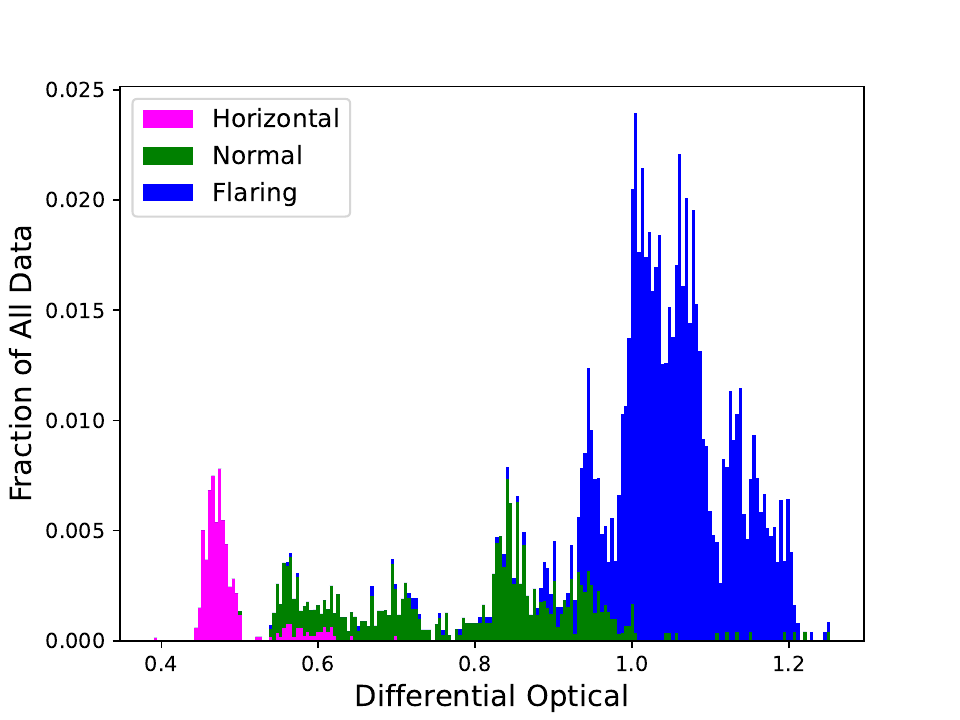}
    \caption{A histogram of the differential optical, divided by the median of the entire differential optical lightcurve.}
    \label{fig:scox1_opt_hist}
\end{figure}

Fig. \ref{fig:scox1_opt_hist} shows a histogram of the optical flux, and helps to illuminate that the step transitions occur at the apexes, at slightly smaller values of $s_z$ than their strict definitions of $s_z$=1 and $s_z$=2.  While its behavior is not what one would generally describe as bimodal, there are two peaks that are worth pointing out.  The first is a low scatter, low intensity peak corresponding to the HB, and the second a broad, high intensity peak corresponding to the FB.  A small peak can be seen at an intensity of $\sim$0.58 (the location of the NB step).  However, the peak frequency is of a similar scale as certain portions of the soft apex, and so is likely too small to call this an example of trimodal optical behavior.  Again, this may be more of an issue with simply not having much solidly NB data, because the step in Fig. \ref{fig:scox1_st_optical} clearly defines the region.  It may be tempting to define a smooth envelope that encompasses the optical-$s_z$ shape in Fig. \ref{fig:scox1_st_optical} instead of describing it as steps.  The regions that are devoid of data (around the jumps to new flux levels in the step description) would then be due to a lack of data coverage.  However, this is unlikely to be an adequate explanation.  Although Fig. \ref{fig:scox1_opt_hist} cannot be described as bimodal (or trimodal), these behaviors have been well recorded in Sco X-1 since not long after its discovery \citep{1967ApJ...150..851H,1970A&A.....8....1H} to more recent works \citep{2016MNRAS.459.3596H}.  This is at odds with an envelope explanation, which would fill in these data gaps.  Treating the behavior like steps will naturally lead to a double (or triple) peaked histogram with enough coverage of each branch.

The $u_z$ variable plotted against others can be seen in Fig. \ref{fig:scox1_uz}. As one would hope for with good initial rank definitions, $t_z$ is more or less uniform for all angles (as are the other variables), albeit with a couple of bulges at $u_z\simeq22.5^\circ$ and $u_z\simeq202.5^\circ$.  The sources of these are the upper FB region where the ``cylinder'' has squished along one axis, as well as a region along the soft apex.  With the inclusion of intensity into the ($s_z$,$t_z$,$u_z$) mapping, the lack of any $s_z$ correlation with $t_z$ or $u_z$ leads to the conclusion that $s_z$ is the dominant parameter when describing the spectral state of Sco X-1.

\section{Discussion}
\subsection{Potential Reprocessing}
Fig. \ref{fig:ccd_blips} plots the X-ray data corresponding to the minor peaks as larger, coloured points plotted over the full CD.  The figure shows that minor peaks were only identified on the FB and at the soft apex.  This lends further evidence to the minor peaks being reprocessing peaks, as reprocessing has only ever been found on or near the FB.  The interpretation for this depends on which spectral model is used.  For example, in the \citet{1995ApJ...454L.137P} model, the binary begins to move into super-Eddington luminosities on the FB.  The accretion rate becomes tumultuous, radially moving inward in some regions and outward in others.  As material builds up in the outer corona, the electron scattering opacity increases, which causes more X-rays to be redirected onto the accretion disc and reprocessed \citep{2003AJ....125.1437M}.  Independently, the \citet{2012A&A...546A..35C} model says that Sco-likes have increased accretion rates going up the FB.  As much of the X-ray radiation on the FB originates as flares from unstable burning near the NS surface in this model, an increased accretion rate would predict more reprocessing in this case as well.

Out of the nine nights that contained both optical and X-ray data, four had instances of potential reprocessing.  All of the minor peaks occurred during times that contained either flickering, flaring, or some combination of the two in both the X-ray and optical lightcurves.  There are no outliers to this, i.e., no overlapping lightcurves with the aforementioned traits without a minor peak, with one possible exception in Fig. \ref{fig:may19c_full} (which also lies in the FB).  There is clearly flickering in both lightcurves, but the X-ray lightcurve is so short that any minor peaks may have been masked by edge effects.  This leads to the conclusion that FB correlations are always present when there is suitable variability and enough data.

Our conclusions from studying correlated variability are consistent with expectations based on energetics and spectral energy distribution modelling. It has long been appreciated that optical emission from luminous LMXBs should be dominated by reprocessed X-rays. \citet{1994A&A...290..133V} showed that the absolute magnitudes of a sample of 18 LMXBs, including Sco X-1, were consistent with predictions based on their X-ray luminosities and orbital periods (and hence inferred disk sizes). \citet{1991ApJ...376..278V} showed that International Ultraviolet Explorer UV spectra of Sco X-1 were best fitted by an irradiated disk model. More recently, \citet{2007MNRAS.379.1108R} studied X-ray and optical/infrared (OIR) emissions and compared them to what would be expected for OIR dominated by reprocessing, viscous heating, and synchrotron. The authors found that the OIR from Z sources can be explained by X-ray reprocessing alone. Z sources generally having larger amounts of OIR due to reprocessing is consistent with their higher X-ray luminosities and larger disk radii.

It has been shown in \citet{1999MNRAS.303..139D} that the disc is likely convex and self-shielding, but there have been suggestions for how the outer disc could be illuminated.  \citet{2016MNRAS.459.3596H} proposes that Sco X-1 undergoes obscurations in the inner disc, and that the amount of outer disc illumination depends on changes in this obscuration.  For example, a model similar to the one in \citet{2014ApJ...789...98T}, with a central obscuring torus, would scatter soft X-rays originating around the NS, and prevent them from reaching the outer disc.  A Comptonizing accretion disc corona, like the one proposed in \citet{2012A&A...546A..35C}, could obscure similarly.  Decreasing the thickness of the torus, or its optical depth, would allow more soft photons to reach the outer accretion disc.

The outer disc could be further illuminated with the addition of disc warping.  In this process, uneven illumination on each side of the outer disc causes a torque on the annulus of unequal irradiation.  As material moves to the inner radii, it carries with it a ``memory'' of said torque.  This results in a warp with nodes in the shape of a prograde spiral, rotating with a super-orbital periodicity.  Sco X-1 is known to be ``indeterminately unstable'' to disc warping, experiencing times where the disc is warped and times where the disc is flat \citep{2001MNRAS.320..485O}.  However, as \citet{2012MNRAS.420.1575K} notes, its 62 hour super-orbital period \citep{1989PAZh...15.1072K} is not always a strong effect within the lightcurve, and so it is likely that warping is not significantly influencing disc irradiation here.

\subsection{Behaviors on the Z Track}
\citet{2015MNRAS.451.3857S} found that optical data could be used to predict whether Sco X-1 was on the NB or FB when no X-ray spectral data was available.  The results in Fig. \ref{fig:scox1_opt_hist} back that claim up and expand upon it by including the HB.  They also say that their results are not perfect discriminators of the two branches, noting long tails in the optical bimodal behavior.  Our results are more promising, showing that when using an $s_z$ parameterization, the muddiness disappears somewhat.  The soft apex appears here as a clear transition region between the NB and the FB, containing only intermediate fluxes between the two steps.

The model in \citet{1995ApJ...454L.137P} can explain this type of optical behavior, as interpreted by \citet{2003AJ....125.1437M}.  In the HB, the release of kinetic energy from infalling material would cause a temperature increase in the magnetosphere and central corona.  This would lead to an increase in both the optical and X-ray intensities, with both being tied together based on the accretion rate (amount of infalling material/released kinetic energy).  In Fig. \ref{fig:hid}, one can see that the X-ray intensity is more or less constant in the HB, and so this model would predict similar behavior in the optical (confirmed by the HB step in Fig. \ref{fig:scox1_st_optical}).  On the normal branch, the radiation pressure increases, leading to a pileup of material near the neutron star surface.  This causes an increase in the amount of X-rays that are absorbed and re-emitted as optical light, so the NB step would be expected to be higher than the HB step, even as X-ray intensity decreases.  As the Eddington critical rate is passed and the mass flow becomes chaotic (FB), material builds up in the corona, leading to more X-rays being scattered down to the outer accretion disc for reprocessing.  This would lead to another jump in the optical intensity, correlating with X-ray intensity.  This would also explain the additional optical scatter in the FB, as the X-rays on that branch behave similarly.

The point of transition from uncorrelated to correlated is notable, as other papers have noted a ``kink'' occurring there on the FB \citep{2005ApJ...623.1070M}.  The kink location is where the FB slope increases, between $s_z=4.0$ and $s_z=4.5$, and can be seen more clearly in the CDs of \citet{1992ApJ...396..201H} and \citet{2000MNRAS.311..201D}.  \citet{2003A&A...405..237B}, who use a spectral model where the neutron star is the blackbody emitter, describe this hard colour increase as the point where the emission covers the entire central object.  In addition, \citet{2005ApJ...623.1070M} found that the FB after the kink could be identified with only optical data, using Johnson B magnitude and the presence of the largest amplitude (0.1 mag) flares within 8-minute data segments.  This adds credence to the results in Fig. \ref{fig:scox1_st_optical}.

\citet{2003AJ....125.1437M} discuss the speeds at which Sco X-1 moves through its Z track, which the authors say may need more data for further confirmation.  They point out that the system begins to slow down as it moves down the FB, to a minimum at the soft apex, before speeding up again approaching the hard apex, where it reaches a local maximum.  It then slowly decreases up the HB.  Our data do not quite back every aspect of this up.  Although there is agreement in that the speed decreases from the top of the FB, this change only lasts for the upper half.  From there, the maximum speed decreases slightly as the system approaches the soft apex, where it seems to remain until the hard apex (the mapping artifact makes it difficult to tell).  There is a small, but noticeable increase in speeds at the hard apex, but it does not decrease through the HB.  \citet{1992ApJ...396..201H} also notes a taper at the Sco X-1 soft apex, and point out that since $ds_z/dt$ is a function of $s_z$, the shortest timescale for intensity fluctuations caused by super-Eddington instabilities will be dependent on the accretion rate.  This should hold for our data as well, at least in the upper FB.

\section{Conclusions}
We have analyzed simultaneous 1 s Argos optical data and RXTE X-ray data for the Z source Sco X-1.  Discrete cross correlations of overlapping sets of these data revealed two CCF phenomena:
\begin{itemize}
    \item One larger potential reprocessing peak that stood out from the surrounding CCF features.  Its peak maximum occurred at $\sim$1 s.
    \item Eight smaller minor peaks that were much more difficult to discern from the other CCF features.  Their peaks all occur at lags less than 4 s.
\end{itemize}
Due to the similarity of a stacked version of the minor peaks to the larger peak, it is likely that they are just weaker versions of the same.  To confirm that the minor peaks were in fact real, a scheme was created so that they could be identified computationally, using a series of criteria involving peak properties and high-pass Butterworth filters.  The same identification scheme was tested on the CCFs of sections of data that did not overlap in time.  Poisson statistics performed on the results led to the conclusion that the minor peaks are occurring at a statistically significant rate (the chances of even one minor peak being a false positive is calculated as $\sim$7\%).  Further evidence comes from the CD, as all of the minor peaks occurred on the FB or soft apex.  No other potential reprocessing events were seen outside of these Z track regions. Comparing minor peak lags to the expected companion lag and phase leads to the implication that the disc would be the reprocessing region.  However, companion reprocessing could still make some contribution to the tails of the minor peaks.  Additional checking of the moments of the peaks yielded no new information, except confirming slower decay timescales (in skewness).  Ultimately, the results support the conclusion that short lag FB correlations are always present, so long as there is an adequate amount of data and suitable variability.

In addition, it was found that the Z track location could be estimated from the optical intensity alone.  We performed an analysis of the Z track behavior by using a 3D modification of the rank number scheme (Appendix \ref{sz_apx}), which when plotted against the optical can be divided into four distinct regions:
\begin{itemize}
    \item $s_z < 1$: This region contains no optical-$s_z$ correlation, and occurs at a differential optical flux of about $\sim$0.47.  This corresponds to the HB.
    \item $1 < s_z < 2$: This region contains no optical-$s_z$ correlation, and occurs at a differential optical flux of about $\sim$0.58.  This corresponds to the NB.
    \item $2 < s_z < 4$: This region contains no optical-$s_z$ correlation, and occurs at a differential optical flux of about $\sim$1.1.  This corresponds to the lower half of the FB.
    \item $s_z > 4$: This region contains a positive optical-$s_z$ correlation.  This corresponds to the upper half of the FB.
\end{itemize}
The division at $s_z = 4$ is notable, as it is associated with a kink in the FB that is interpreted as the point where blackbody emission covers the entire central object.  A histogram of the optical intensity reveals that the soft apex is a clear transition region with only intermediate fluxes between the NB and FB steps.

\section*{Acknowledgements}
A. B. I. acknowledges support from a Louisiana Board of Regents Fellowship and a Graduate Student Research Assistance award from the Louisiana Space Grant Consortium. This work was also supported by NASA/Louisiana Board of Regents grant NNX07AT62A/LEQSF(2007-10) Phase3-02. This research has made use of NASA’s Astrophysics Data System.  This paper includes data taken at The McDonald Observatory of The University of Texas at Austin.  This research has made use of data and software provided by the High Energy Astrophysics Science Archive Research Center (HEASARC), which is a service of the Astrophysics Science Division at NASA's Goddard Space Flight Center, as well as NASA’s Astrophysics Data System. We would like to thank the Rossi X-ray Timing Explorer team for their support and especially the schedulers for facilitating simultaneous observations.  Finally, we thank the referee for the helpful comments they provided.

%%%%%%%%%%%%%%%%%%%%%%%%%%%%%%%%%%%%%%%%%%%%%%%%%%
\section*{Data Availability}
The X-ray data used in this study can be found on HEASARC, using the observation ID 94318.  The optical data are available on Zenodo (DOI: 10.5281/zenodo.8250724).



\bibliographystyle{mnras}
\typeout{}
\bibliography{biblio} % if your bibtex file is called example.bib

\begin{thebibliography}{}
\makeatletter
\relax
\def\mn@urlcharsother{\let\do\@makeother \do\$\do\&\do\#\do\^\do\_\do\%\do\~}
\def\mn@doi{\begingroup\mn@urlcharsother \@ifnextchar [ {\mn@doi@}
  {\mn@doi@[]}}
\def\mn@doi@[#1]#2{\def\@tempa{#1}\ifx\@tempa\@empty \href
  {http://dx.doi.org/#2} {doi:#2}\else \href {http://dx.doi.org/#2} {#1}\fi
  \endgroup}
\def\mn@eprint#1#2{\mn@eprint@#1:#2::\@nil}
\def\mn@eprint@arXiv#1{\href {http://arxiv.org/abs/#1} {{\tt arXiv:#1}}}
\def\mn@eprint@dblp#1{\href {http://dblp.uni-trier.de/rec/bibtex/#1.xml}
  {dblp:#1}}
\def\mn@eprint@#1:#2:#3:#4\@nil{\def\@tempa {#1}\def\@tempb {#2}\def\@tempc
  {#3}\ifx \@tempc \@empty \let \@tempc \@tempb \let \@tempb \@tempa \fi \ifx
  \@tempb \@empty \def\@tempb {arXiv}\fi \@ifundefined
  {mn@eprint@\@tempb}{\@tempb:\@tempc}{\expandafter \expandafter \csname
  mn@eprint@\@tempb\endcsname \expandafter{\@tempc}}}

\bibitem[\protect\citeauthoryear{{Augusteijn} et~al.,}{{Augusteijn}
  et~al.}{1992}]{1992A&A...265..177A}
{Augusteijn} T.,  et~al., 1992, \aap, \href
  {https://ui.adsabs.harvard.edu/abs/1992A&A...265..177A} {265, 177}

\bibitem[\protect\citeauthoryear{{Barnard}, {Church}  \&
  {Ba{\l}uci{\'n}ska-Church}}{{Barnard} et~al.}{2003}]{2003A&A...405..237B}
{Barnard} R.,  {Church} M.~J.,   {Ba{\l}uci{\'n}ska-Church} M.,  2003, \mn@doi
  [\aap] {10.1051/0004-6361:20030539}, \href
  {https://ui.adsabs.harvard.edu/abs/2003A&A...405..237B} {405, 237}

\bibitem[\protect\citeauthoryear{{Bradshaw}, {Fomalont}  \&
  {Geldzahler}}{{Bradshaw} et~al.}{1999}]{1999ApJ...512L.121B}
{Bradshaw} C.~F.,  {Fomalont} E.~B.,   {Geldzahler} B.~J.,  1999, \mn@doi
  [\apjl] {10.1086/311889}, \href
  {https://ui.adsabs.harvard.edu/abs/1999ApJ...512L.121B} {512, L121}

\bibitem[\protect\citeauthoryear{Butterworth}{Butterworth}{1930}]{Butterworth1930}
Butterworth S.,  1930, Experimental Wireless \& the Wireless Engineer, 7, 536

\bibitem[\protect\citeauthoryear{{Cherepashchuk}, {Khruzina}  \&
  {Bogomazov}}{{Cherepashchuk} et~al.}{2021}]{2021MNRAS.508.1389C}
{Cherepashchuk} A.~M.,  {Khruzina} T.~S.,   {Bogomazov} A.~I.,  2021, \mn@doi
  [\mnras] {10.1093/mnras/stab2515}, \href
  {https://ui.adsabs.harvard.edu/abs/2021MNRAS.508.1389C} {508, 1389}

\bibitem[\protect\citeauthoryear{{Cherepashchuk}, {Khruzina}  \&
  {Bogomazov}}{{Cherepashchuk} et~al.}{2022}]{2022ARep...66..348C}
{Cherepashchuk} A.~M.,  {Khruzina} T.~S.,   {Bogomazov} A.~I.,  2022, \mn@doi
  [Astronomy Reports] {10.1134/S1063772922040023}, \href
  {https://ui.adsabs.harvard.edu/abs/2022ARep...66..348C} {66, 348}

\bibitem[\protect\citeauthoryear{{Church}, {Gibiec}, {Ba{\l}uci{\'n}ska-Church}
   \& {Jackson}}{{Church} et~al.}{2012}]{2012A&A...546A..35C}
{Church} M.~J.,  {Gibiec} A.,  {Ba{\l}uci{\'n}ska-Church} M.,   {Jackson}
  N.~K.,  2012, \mn@doi [\aap] {10.1051/0004-6361/201218987}, \href
  {https://ui.adsabs.harvard.edu/abs/2012A&A...546A..35C} {546, A35}

\bibitem[\protect\citeauthoryear{{Cominsky}, {London}  \& {Klein}}{{Cominsky}
  et~al.}{1987}]{1987ApJ...315..162C}
{Cominsky} L.~R.,  {London} R.~A.,   {Klein} R.~I.,  1987, \mn@doi [\apj]
  {10.1086/165122}, \href
  {https://ui.adsabs.harvard.edu/abs/1987ApJ...315..162C} {315, 162}

\bibitem[\protect\citeauthoryear{{Cowley} \& {Crampton}}{{Cowley} \&
  {Crampton}}{1975}]{1975ApJ...201L..65C}
{Cowley} A.~P.,  {Crampton} D.,  1975, \mn@doi [\apjl] {10.1086/181943}, \href
  {https://ui.adsabs.harvard.edu/abs/1975ApJ...201L..65C} {201, L65}

\bibitem[\protect\citeauthoryear{{Dieters} \& {van der Klis}}{{Dieters} \& {van
  der Klis}}{2000}]{2000MNRAS.311..201D}
{Dieters} S.~W.,  {van der Klis} M.,  2000, \mn@doi [\mnras]
  {10.1046/j.1365-8711.2000.03050.x}, \href
  {https://ui.adsabs.harvard.edu/abs/2000MNRAS.311..201D} {311, 201}

\bibitem[\protect\citeauthoryear{{Dubus}, {Lasota}, {Hameury}  \&
  {Charles}}{{Dubus} et~al.}{1999}]{1999MNRAS.303..139D}
{Dubus} G.,  {Lasota} J.-P.,  {Hameury} J.-M.,   {Charles} P.,  1999, \mn@doi
  [\mnras] {10.1046/j.1365-8711.1999.02212.x}, \href
  {https://ui.adsabs.harvard.edu/abs/1999MNRAS.303..139D} {303, 139}

\bibitem[\protect\citeauthoryear{{Edelson} \& {Krolik}}{{Edelson} \&
  {Krolik}}{1988}]{1988ApJ...333..646E}
{Edelson} R.~A.,  {Krolik} J.~H.,  1988, \mn@doi [\apj] {10.1086/166773}, \href
  {https://ui.adsabs.harvard.edu/abs/1988ApJ...333..646E} {333, 646}

\bibitem[\protect\citeauthoryear{{Fomalont}, {Geldzahler}  \&
  {Bradshaw}}{{Fomalont} et~al.}{2001}]{2001ApJ...558..283F}
{Fomalont} E.~B.,  {Geldzahler} B.~J.,   {Bradshaw} C.~F.,  2001, \mn@doi
  [\apj] {10.1086/322479}, \href
  {https://ui.adsabs.harvard.edu/abs/2001ApJ...558..283F} {558, 283}

\bibitem[\protect\citeauthoryear{Frenet}{Frenet}{1852}]{Frenet1852}
Frenet F.,  1852, Journal de Mathématiques Pures et Appliquées, pp 437--447

\bibitem[\protect\citeauthoryear{{Gaia Collaboration} et~al.,}{{Gaia
  Collaboration} et~al.}{2021}]{refId0}
{Gaia Collaboration} et~al., 2021, \mn@doi [A\&A]
  {10.1051/0004-6361/202039657}, 649, A1

\bibitem[\protect\citeauthoryear{{Galloway}, {Muno}, {Hartman}, {Psaltis}  \&
  {Chakrabarty}}{{Galloway} et~al.}{2008}]{2008ApJS..179..360G}
{Galloway} D.~K.,  {Muno} M.~P.,  {Hartman} J.~M.,  {Psaltis} D.,
  {Chakrabarty} D.,  2008, \mn@doi [\apjs] {10.1086/592044}, \href
  {https://ui.adsabs.harvard.edu/abs/2008ApJS..179..360G} {179, 360}

\bibitem[\protect\citeauthoryear{{Galloway}, {Premachandra}, {Steeghs},
  {Marsh}, {Casares}  \& {Cornelisse}}{{Galloway}
  et~al.}{2014}]{2014ApJ...781...14G}
{Galloway} D.~K.,  {Premachandra} S.,  {Steeghs} D.,  {Marsh} T.,  {Casares}
  J.,   {Cornelisse} R.,  2014, \mn@doi [\apj] {10.1088/0004-637X/781/1/14},
  \href {https://ui.adsabs.harvard.edu/abs/2014ApJ...781...14G} {781, 14}

\bibitem[\protect\citeauthoryear{{Gaskell} \& {Peterson}}{{Gaskell} \&
  {Peterson}}{1987}]{1987ApJS...65....1G}
{Gaskell} C.~M.,  {Peterson} B.~M.,  1987, \mn@doi [\apjs] {10.1086/191216},
  \href {https://ui.adsabs.harvard.edu/abs/1987ApJS...65....1G} {65, 1}

\bibitem[\protect\citeauthoryear{{Giacconi}, {Gursky}, {Paolini}  \&
  {Rossi}}{{Giacconi} et~al.}{1962}]{1962PhRvL...9..439G}
{Giacconi} R.,  {Gursky} H.,  {Paolini} F.~R.,   {Rossi} B.~B.,  1962, \mn@doi
  [\prl] {10.1103/PhysRevLett.9.439}, \href
  {https://ui.adsabs.harvard.edu/abs/1962PhRvL...9..439G} {9, 439}

\bibitem[\protect\citeauthoryear{{Gottlieb}, {Wright}  \& {Liller}}{{Gottlieb}
  et~al.}{1975}]{1975ApJ...195L..33G}
{Gottlieb} E.~W.,  {Wright} E.~L.,   {Liller} W.,  1975, \mn@doi [\apjl]
  {10.1086/181703}, \href
  {https://ui.adsabs.harvard.edu/abs/1975ApJ...195L..33G} {195, L33}

\bibitem[\protect\citeauthoryear{{Hasinger} \& {van der Klis}}{{Hasinger} \&
  {van der Klis}}{1989}]{1989A&A...225...79H}
{Hasinger} G.,  {van der Klis} M.,  1989, \aap, \href
  {https://ui.adsabs.harvard.edu/abs/1989A&A...225...79H} {225, 79}

\bibitem[\protect\citeauthoryear{{Hertz}, {Vaughan}, {Wood}, {Norris},
  {Mitsuda}, {Michelson}  \& {Dotani}}{{Hertz}
  et~al.}{1992}]{1992ApJ...396..201H}
{Hertz} P.,  {Vaughan} B.,  {Wood} K.~S.,  {Norris} J.~P.,  {Mitsuda} K.,
  {Michelson} P.~F.,   {Dotani} T.,  1992, \mn@doi [\apj] {10.1086/171710},
  \href {https://ui.adsabs.harvard.edu/abs/1992ApJ...396..201H} {396, 201}

\bibitem[\protect\citeauthoryear{{Hiltner} \& {Mook}}{{Hiltner} \&
  {Mook}}{1967}]{1967ApJ...150..851H}
{Hiltner} W.~A.,  {Mook} D.~E.,  1967, \mn@doi [\apj] {10.1086/149388}, \href
  {https://ui.adsabs.harvard.edu/abs/1967ApJ...150..851H} {150, 851}

\bibitem[\protect\citeauthoryear{{Hiltner} \& {Mook}}{{Hiltner} \&
  {Mook}}{1970}]{1970A&A.....8....1H}
{Hiltner} W.~A.,  {Mook} D.~E.,  1970, \aap, \href
  {https://ui.adsabs.harvard.edu/abs/1970A&A.....8....1H} {8, 1}

\bibitem[\protect\citeauthoryear{{Homan} et~al.,}{{Homan}
  et~al.}{2007a}]{2007ApJ...656..420H}
{Homan} J.,  et~al., 2007a, \mn@doi [\apj] {10.1086/510447}, \href
  {https://ui.adsabs.harvard.edu/abs/2007ApJ...656..420H} {656, 420}

\bibitem[\protect\citeauthoryear{{Homan}, {Belloni}, {Wijnands}, {van der
  Klis}, {Swank}, {Smith}, {Pereira}  \& {Markwardt}}{{Homan}
  et~al.}{2007b}]{2007ATel.1144....1H}
{Homan} J.,  {Belloni} T.,  {Wijnands} R.,  {van der Klis} M.,  {Swank} J.,
  {Smith} E.,  {Pereira} D.,   {Markwardt} C.,  2007b, The Astronomer's
  Telegram, \href {https://ui.adsabs.harvard.edu/abs/2007ATel.1144....1H}
  {1144, 1}

\bibitem[\protect\citeauthoryear{{Hynes} \& {Britt}}{{Hynes} \&
  {Britt}}{2012}]{2012ApJ...755...66H}
{Hynes} R.~I.,  {Britt} C.~T.,  2012, \mn@doi [\apj]
  {10.1088/0004-637X/755/1/66}, \href
  {https://ui.adsabs.harvard.edu/abs/2012ApJ...755...66H} {755, 66}

\bibitem[\protect\citeauthoryear{{Hynes} et~al.,}{{Hynes}
  et~al.}{2003}]{2003MNRAS.345..292H}
{Hynes} R.~I.,  et~al., 2003, \mn@doi [\mnras]
  {10.1046/j.1365-8711.2003.06938.x}, \href
  {https://ui.adsabs.harvard.edu/abs/2003MNRAS.345..292H} {345, 292}

\bibitem[\protect\citeauthoryear{{Hynes}, {Brien}, {Mullally}  \&
  {Ashcraft}}{{Hynes} et~al.}{2009}]{2009MNRAS.399..281H}
{Hynes} R.~I.,  {Brien} K.~O.,  {Mullally} F.,   {Ashcraft} T.,  2009, \mn@doi
  [\mnras] {10.1111/j.1365-2966.2009.15260.x}, \href
  {https://ui.adsabs.harvard.edu/abs/2009MNRAS.399..281H} {399, 281}

\bibitem[\protect\citeauthoryear{{Hynes}, {Schaefer}, {Baum}, {Hsu}, {Cherry}
  \& {Scaringi}}{{Hynes} et~al.}{2016}]{2016MNRAS.459.3596H}
{Hynes} R.~I.,  {Schaefer} B.~E.,  {Baum} Z.~A.,  {Hsu} C.-C.,  {Cherry} M.~L.,
    {Scaringi} S.,  2016, \mn@doi [\mnras] {10.1093/mnras/stw854}, \href
  {https://ui.adsabs.harvard.edu/abs/2016MNRAS.459.3596H} {459, 3596}

\bibitem[\protect\citeauthoryear{{Kotze} \& {Charles}}{{Kotze} \&
  {Charles}}{2012}]{2012MNRAS.420.1575K}
{Kotze} M.~M.,  {Charles} P.~A.,  2012, \mn@doi [\mnras]
  {10.1111/j.1365-2966.2011.20146.x}, \href
  {https://ui.adsabs.harvard.edu/abs/2012MNRAS.420.1575K} {420, 1575}

\bibitem[\protect\citeauthoryear{{Kudryavtsev}, {Mamontova}, {Svertilov}  \&
  {Tolstaya}}{{Kudryavtsev} et~al.}{1989}]{1989PAZh...15.1072K}
{Kudryavtsev} M.~I.,  {Mamontova} N.~A.,  {Svertilov} S.~I.,   {Tolstaya}
  E.~D.,  1989, Pisma v Astronomicheskii Zhurnal, \href
  {https://ui.adsabs.harvard.edu/abs/1989PAZh...15.1072K} {15, 1072}

\bibitem[\protect\citeauthoryear{{Kuulkers}, {van der Klis}, {Oosterbroek},
  {Asai}, {Dotani}, {van Paradijs}  \& {Lewin}}{{Kuulkers}
  et~al.}{1994}]{1994A&A...289..795K}
{Kuulkers} E.,  {van der Klis} M.,  {Oosterbroek} T.,  {Asai} K.,  {Dotani} T.,
   {van Paradijs} J.,   {Lewin} W.~H.~G.,  1994, \aap, \href
  {https://ui.adsabs.harvard.edu/abs/1994A&A...289..795K} {289, 795}

\bibitem[\protect\citeauthoryear{{Kuulkers}, {van der Klis}  \&
  {Vaughan}}{{Kuulkers} et~al.}{1996}]{1996A&A...311..197K}
{Kuulkers} E.,  {van der Klis} M.,   {Vaughan} B.~A.,  1996, \aap, \href
  {https://ui.adsabs.harvard.edu/abs/1996A&A...311..197K} {311, 197}

\bibitem[\protect\citeauthoryear{{Lin}, {Remillard}  \& {Homan}}{{Lin}
  et~al.}{2010}]{2010ApJ...719.1350L}
{Lin} D.,  {Remillard} R.~A.,   {Homan} J.,  2010, \mn@doi [\apj]
  {10.1088/0004-637X/719/2/1350}, \href
  {https://ui.adsabs.harvard.edu/abs/2010ApJ...719.1350L} {719, 1350}

\bibitem[\protect\citeauthoryear{{Mata Sanchez}, {Munoz-Darias}, {Casares},
  {Steeghs}, {Ramos Almeida}  \& {Acosta Pulido}}{{Mata Sanchez}
  et~al.}{2015}]{2015MNRAS.449L...1M}
{Mata Sanchez} D.,  {Munoz-Darias} T.,  {Casares} J.,  {Steeghs} D.,  {Ramos
  Almeida} C.,   {Acosta Pulido} J.~A.,  2015, \mn@doi [\mnras]
  {10.1093/mnrasl/slv002}, \href
  {https://ui.adsabs.harvard.edu/abs/2015MNRAS.449L...1M} {449, L1}

\bibitem[\protect\citeauthoryear{{McGowan}, {Charles}, {O'Donoghue}  \&
  {Smale}}{{McGowan} et~al.}{2003}]{2003MNRAS.345.1039M}
{McGowan} K.~E.,  {Charles} P.~A.,  {O'Donoghue} D.,   {Smale} A.~P.,  2003,
  \mn@doi [\mnras] {10.1046/j.1365-8711.2003.07029.x}, \href
  {https://ui.adsabs.harvard.edu/abs/2003MNRAS.345.1039M} {345, 1039}

\bibitem[\protect\citeauthoryear{{McNamara} et~al.,}{{McNamara}
  et~al.}{2003}]{2003AJ....125.1437M}
{McNamara} B.~J.,  et~al., 2003, \mn@doi [\aj] {10.1086/367791}, \href
  {https://ui.adsabs.harvard.edu/abs/2003AJ....125.1437M} {125, 1437}

\bibitem[\protect\citeauthoryear{{McNamara}, {Norwood}, {Harrison}, {Holtzman},
  {Dukes}  \& {Barker}}{{McNamara} et~al.}{2005}]{2005ApJ...623.1070M}
{McNamara} B.~J.,  {Norwood} J.,  {Harrison} T.~E.,  {Holtzman} J.,  {Dukes}
  R.,   {Barker} T.,  2005, \mn@doi [\apj] {10.1086/428640}, \href
  {https://ui.adsabs.harvard.edu/abs/2005ApJ...623.1070M} {623, 1070}

\bibitem[\protect\citeauthoryear{{Mu{\~n}oz-Darias}, {Mart{\'\i}nez-Pais},
  {Casares}, {Dhillon}, {Marsh}, {Cornelisse}, {Steeghs}  \&
  {Charles}}{{Mu{\~n}oz-Darias} et~al.}{2007}]{2007MNRAS.379.1637M}
{Mu{\~n}oz-Darias} T.,  {Mart{\'\i}nez-Pais} I.~G.,  {Casares} J.,  {Dhillon}
  V.~S.,  {Marsh} T.~R.,  {Cornelisse} R.,  {Steeghs} D.,   {Charles} P.~A.,
  2007, \mn@doi [\mnras] {10.1111/j.1365-2966.2007.12045.x}, \href
  {https://ui.adsabs.harvard.edu/abs/2007MNRAS.379.1637M} {379, 1637}

\bibitem[\protect\citeauthoryear{{Mu{\~n}oz-Darias} et~al.,}{{Mu{\~n}oz-Darias}
  et~al.}{2008}]{2008AIPC..984...15M}
{Mu{\~n}oz-Darias} T.,  et~al., 2008, in {Phelan} D.,  {Ryan} O.,   {Shearer}
  A.,  eds,  American Institute of Physics Conference Series Vol. 984, High
  Time Resolution Astrophysics: The Universe at Sub-Second Timescales. pp
  15--22 (\mn@eprint {arXiv} {0709.3500}), \mn@doi{10.1063/1.2896925}

\bibitem[\protect\citeauthoryear{O'Brien}{O'Brien}{2000}]{obrien_2000}
O'Brien K.,  2000, PhD thesis, University of St. Andrews

\bibitem[\protect\citeauthoryear{{O'Brien}, {Horne}, {Hynes}, {Chen}, {Haswell}
   \& {Still}}{{O'Brien} et~al.}{2002}]{2002MNRAS.334..426O}
{O'Brien} K.,  {Horne} K.,  {Hynes} R.~I.,  {Chen} W.,  {Haswell} C.~A.,
  {Still} M.~D.,  2002, \mn@doi [\mnras] {10.1046/j.1365-8711.2002.05530.x},
  \href {https://ui.adsabs.harvard.edu/abs/2002MNRAS.334..426O} {334, 426}

\bibitem[\protect\citeauthoryear{{Ogilvie} \& {Dubus}}{{Ogilvie} \&
  {Dubus}}{2001}]{2001MNRAS.320..485O}
{Ogilvie} G.~I.,  {Dubus} G.,  2001, \mn@doi [\mnras]
  {10.1046/j.1365-8711.2001.04011.x}, \href
  {https://ui.adsabs.harvard.edu/abs/2001MNRAS.320..485O} {320, 485}

\bibitem[\protect\citeauthoryear{{Priedhorsky}, {Hasinger}, {Lewin},
  {Middleditch}, {Parmar}, {Stella}  \& {White}}{{Priedhorsky}
  et~al.}{1986}]{1986ApJ...306L..91P}
{Priedhorsky} W.,  {Hasinger} G.,  {Lewin} W.~H.~G.,  {Middleditch} J.,
  {Parmar} A.,  {Stella} L.,   {White} N.,  1986, \mn@doi [\apjl]
  {10.1086/184712}, \href
  {https://ui.adsabs.harvard.edu/abs/1986ApJ...306L..91P} {306, L91}

\bibitem[\protect\citeauthoryear{{Psaltis}, {Lamb}  \& {Miller}}{{Psaltis}
  et~al.}{1995}]{1995ApJ...454L.137P}
{Psaltis} D.,  {Lamb} F.~K.,   {Miller} G.~S.,  1995, \mn@doi [\apjl]
  {10.1086/309780}, \href
  {https://ui.adsabs.harvard.edu/abs/1995ApJ...454L.137P} {454, L137}

\bibitem[\protect\citeauthoryear{{Remillard}, {Lin}, {ASM Team at MIT}  \&
  {NASA/GSFC}}{{Remillard} et~al.}{2006}]{2006ATel..696....1R}
{Remillard} R.~A.,  {Lin} D.,  {ASM Team at MIT}  {NASA/GSFC} 2006, The
  Astronomer's Telegram, \href
  {https://ui.adsabs.harvard.edu/abs/2006ATel..696....1R} {696, 1}

\bibitem[\protect\citeauthoryear{{Russell}, {Fender}  \& {Jonker}}{{Russell}
  et~al.}{2007}]{2007MNRAS.379.1108R}
{Russell} D.~M.,  {Fender} R.~P.,   {Jonker} P.~G.,  2007, \mn@doi [\mnras]
  {10.1111/j.1365-2966.2007.12008.x}, \href
  {https://ui.adsabs.harvard.edu/abs/2007MNRAS.379.1108R} {379, 1108}

\bibitem[\protect\citeauthoryear{{Sandage} et~al.,}{{Sandage}
  et~al.}{1966}]{1966ApJ...146..316S}
{Sandage} A.,  et~al., 1966, \mn@doi [\apj] {10.1086/148892}, \href
  {https://ui.adsabs.harvard.edu/abs/1966ApJ...146..316S} {146, 316}

\bibitem[\protect\citeauthoryear{{Scaringi}, {Maccarone}, {Hynes},
  {K{\"o}rding}, {Ponti}, {Knigge}, {Britt}  \& {van Winckel}}{{Scaringi}
  et~al.}{2015}]{2015MNRAS.451.3857S}
{Scaringi} S.,  {Maccarone} T.~J.,  {Hynes} R.~I.,  {K{\"o}rding} E.,  {Ponti}
  G.,  {Knigge} C.,  {Britt} C.~T.,   {van Winckel} H.,  2015, \mn@doi [\mnras]
  {10.1093/mnras/stv1216}, \href
  {https://ui.adsabs.harvard.edu/abs/2015MNRAS.451.3857S} {451, 3857}

\bibitem[\protect\citeauthoryear{Serret}{Serret}{1851}]{Serret1851}
Serret J.-A.,  1851, Journal de Mathématiques Pures et Appliquées, pp
  193--207

\bibitem[\protect\citeauthoryear{{Titarchuk}, {Seifina}  \&
  {Shrader}}{{Titarchuk} et~al.}{2014}]{2014ApJ...789...98T}
{Titarchuk} L.,  {Seifina} E.,   {Shrader} C.,  2014, \mn@doi [\apj]
  {10.1088/0004-637X/789/2/98}, \href
  {https://ui.adsabs.harvard.edu/abs/2014ApJ...789...98T} {789, 98}

\bibitem[\protect\citeauthoryear{{Vrtilek}, {Penninx}, {Raymond}, {Verbunt},
  {Hertz}, {Wood}, {Lewin}  \& {Mitsuda}}{{Vrtilek}
  et~al.}{1991}]{1991ApJ...376..278V}
{Vrtilek} S.~D.,  {Penninx} W.,  {Raymond} J.~C.,  {Verbunt} F.,  {Hertz} P.,
  {Wood} K.,  {Lewin} W.~H.~G.,   {Mitsuda} K.,  1991, \mn@doi [\apj]
  {10.1086/170278}, \href
  {https://ui.adsabs.harvard.edu/abs/1991ApJ...376..278V} {376, 278}

\bibitem[\protect\citeauthoryear{{van Paradijs} \& {McClintock}}{{van Paradijs}
  \& {McClintock}}{1994}]{1994A&A...290..133V}
{van Paradijs} J.,  {McClintock} J.~E.,  1994, \aap, \href
  {https://ui.adsabs.harvard.edu/abs/1994A&A...290..133V} {290, 133}

\makeatother
\end{thebibliography}

%\begin{thebibliography}{99}
%\end{thebibliography}

%%%%%%%%%%%%%%%%%%%%%%%%%%%%%%%%%%%%%%%%%%%%%%%%%%

%%%%%%%%%%%%%%%%% APPENDICES %%%%%%%%%%%%%%%%%%%%%

\appendix

\section{Z Track Parameterization}
\label{sz_apx}
\begin{figure}
    \includegraphics[width=\columnwidth]{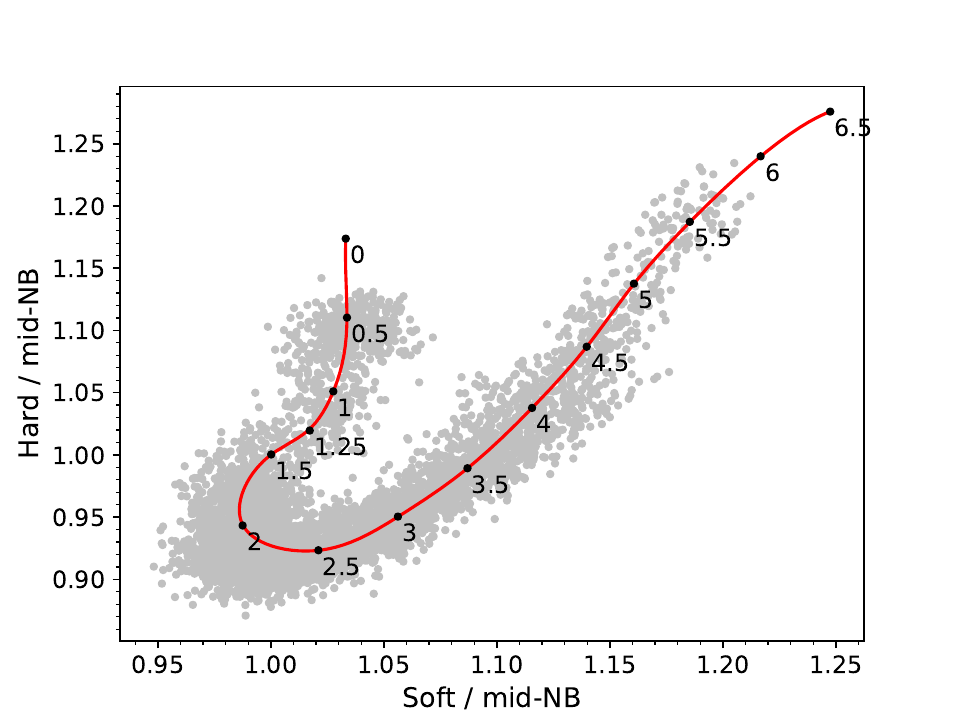}
    \caption{The CD overplotted by the $s_z$ spline and the defined ranks interpolated to get it.  The ranks of 1 and 2 were placed on the hard and soft apexes respectively of a colour-colour-intensity plot.  Ranks between and around those were filled in using that defined scale (the size of the normal branch).}
    \label{fig:ranks}
\end{figure}

To quantify the location on the Z track, a rank number system similar to the one in \citet{2000MNRAS.311..201D} was used.  The method defines the coordinate system ($s_z$,$t_z$), where $s_z$ is the projected location of a data point along the Z track, and $t_z$ is the distance away from the Z track.  Values called ``ranks'' were first assigned by hand to different reference points along the Z track (Fig. \ref{fig:ranks}).  These ranks are the basis for what will become the $s_z$ variable.  The hard apex was defined as rank 1, the soft apex as rank 2, and the intermediate and outer points were filled in using the normalized arc length between the two initial ranks (i.e., if the normalized arc length of the NB was $L$, then an arc length of $s\approx L$ away from the soft apex onto the FB would correspond to a rank of 3).  Thus, the scale of the ranks is dependent on the length of the NB, and there is no defined minimum or maximum value.  Two separate cubic spline interpolations of the ranks with the data along each colour axis results in the Z track used to calculate $s_z$ and $t_z$, seen in Fig. \ref{fig:ranks}.  Relying on the NB arc length for scale allows for the ranks to be as large as necessary to accommodate different lengths of the HB and FB.  It also means that $ds_z/dt$ is directly comparable in different branches, as it represents how quickly $s_z$ changes within a CD.

As intensity is often an important variable in defining branches, this parameterization scheme was expanded further to include the intensity axis.  Moving the spline into three dimensions required performing the aforementioned steps one more time.  This would naturally lead to another parameter, which will from here on out be referred to as $u_z$.  In 3D, $s_z$ and $t_z$ resemble the height and radius respectively of a ``warped'' cylindrical coordinate system.  If one imagines the soft colour, hard colour, and intensity axes as a Cartesian coordinate system, then the warped cylindrical height axis ($s_z$) meanders within, not necessarily a straight line, and not necessarily parallel to one of the Cartesian axes. $u_z$, then, is analogous to the angular component.  While an interesting parameter, most of its utility comes from checking whether or not the defined ranks were well placed.  It should be noted that, because the scale of intensity is so much larger than the scales of the colours, all data were divided by the midpoint of the NB ($s_z$=1.5), so that parameter values were not dominated by any individual axis.

\begin{figure}
    \includegraphics[width=\columnwidth]{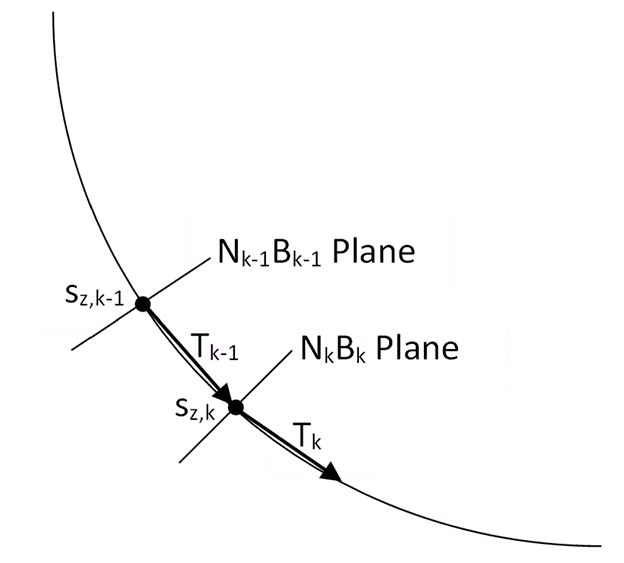}
    \caption{A demonstration of how the $N_{k-1}B_{k-1}$ plane transforms into the $N_kB_K$ plane (distance between the two is exaggerated for clarity) along a small segment of the Z track.  $T_k$ is the vector that points from $s_{z,k-1}$ to $s_{z,k}$.  The directions of $N_k$ and $B_k$ relies on how $T_{k-1}$ is rotated around two separate axes to become parallel with $T_k$.}
    \label{fig:parameterization}
\end{figure}

Returning to the previous analogy, in a standard cylindrical coordinate system, the vectors defining the angular origin are parallel, regardless of the location on the height axis.  In fact, the planes normal to the height axis are all parallel as well, because the height axis is a straight line.  When the height axis is warped, the normal plane must turn with it, and so the set of planes normal to the $s_z$ axis (and the angular origin vectors) are not parallel to each other.  Therefore, calculating $u_z$ requires a system where a set of Cartesian axes rotate along $s_z$.  Borrowing terminology from the similar Frenet-Serret formulae \citep{Frenet1852,Serret1851}, these axes can be defined as the tangent ($T$), the normal ($N$), and the binormal ($B$), and can be calculated inductively.  $T$ is the easiest to calculate, as it always points with $s_z$ (Fig. \ref{fig:parameterization}).  $N$ and $B$ both lie on the plane perpendicular to $T$, and their directions directly determine the origin point of $u_z$.  The initial positions of $N_1$ and $B_1$ are arbitrary, so long as they are orthogonal and lie on the $N_1B_1$ plane.  To find $N_k$ and $B_k$, axes undergo two rotations, designed to make $T_{k-1}$ parallel to $T_k$.  The first was around an axis parallel to the intensity axis, passing through the Z track at the $T_kN_kB_k$ origin, at an angle such that the rotated $T_{k-1}'$ is parallel to $T_k$ when both are projected onto the soft-hard plane.  The second rotation is simply turning $T_{k-1}'$ to the same orientation as $T_k$.  The question could arise as to why the aforementioned Frenet-Serret formulas were not used, as they are applied to similar problems.  In those equations, the $N$ axis is defined as $N=dT/ds$, where $s$ represents the arc length.  Thus, if $T$ were to ``wiggle'', the $N$ and $B$ axes would rotate very quickly.  The system described above adds some stiffness to the rotation of the NB plane, so that the $u_z$ origin will not suddenly move to a different side of the Z track.

\clearpage
\section{Other instances of Optical/X-Ray Overlap}
\label{nonblip_ccf_apx}

In an effort to be consistent with public data availability goals, the following contains all overlapping X-ray and optical data that did not result in a near-zero lag correlation.  The differential optical light curves have been divided by the median of the observation.  The bottom subplot contains the unfiltered CCF (the filtered versions do not reveal any reprocessing peaks).

\begin{figure}
    \begin{subfigure}[b]{.475\textwidth}
        \setcounter{subfigure}{0}
        \includegraphics[width=\textwidth]{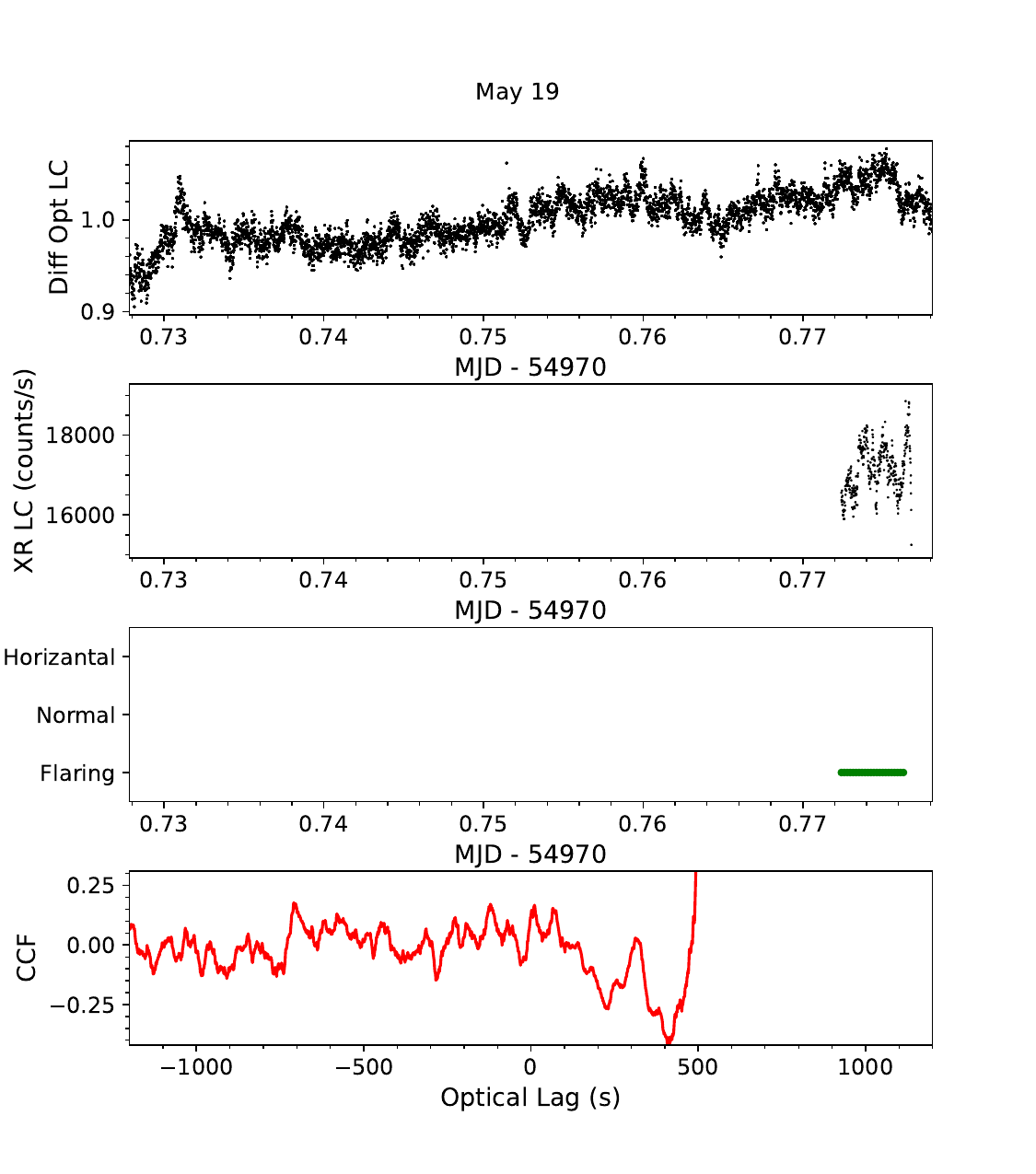}
        \caption[empty]{}
        \label{fig:may19c_full}
    \end{subfigure}
    \hfill
    \begin{subfigure}[b]{.475\textwidth}
        \setcounter{subfigure}{1}
        \includegraphics[width=\textwidth]{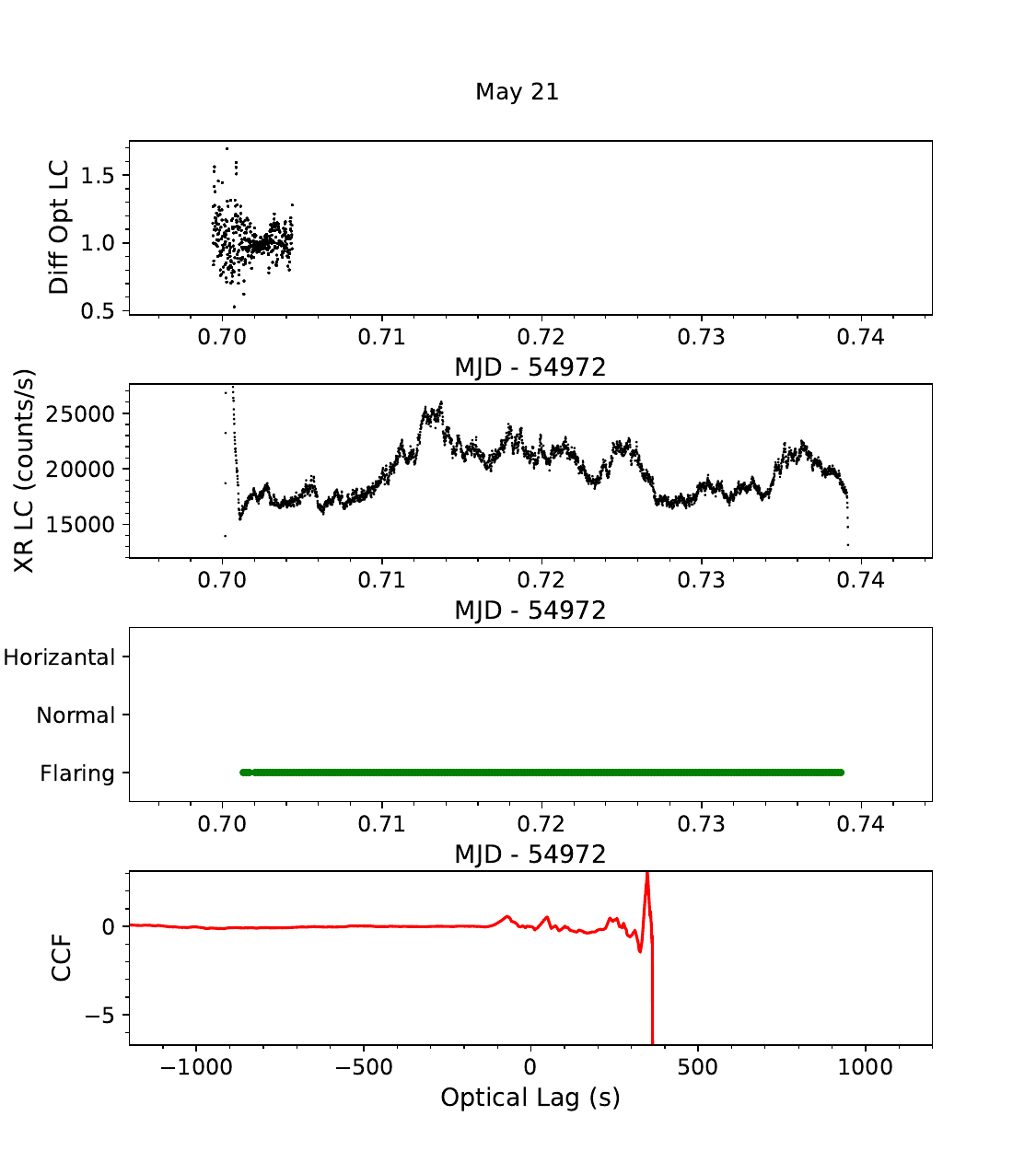}
        \caption[empty]{}
        \label{fig:may21_full}
    \end{subfigure}
    \caption{The first plot is the differential optical lightcurve (normalized by the median), the second is the X-ray intensity, the third is the location on the Z track, and the last shows the unfiltered CCF.}
\end{figure}

\begin{figure*}
    \ContinuedFloat
    \begin{subfigure}[b]{.475\textwidth}
        \setcounter{subfigure}{2}
        \includegraphics[width=\textwidth]{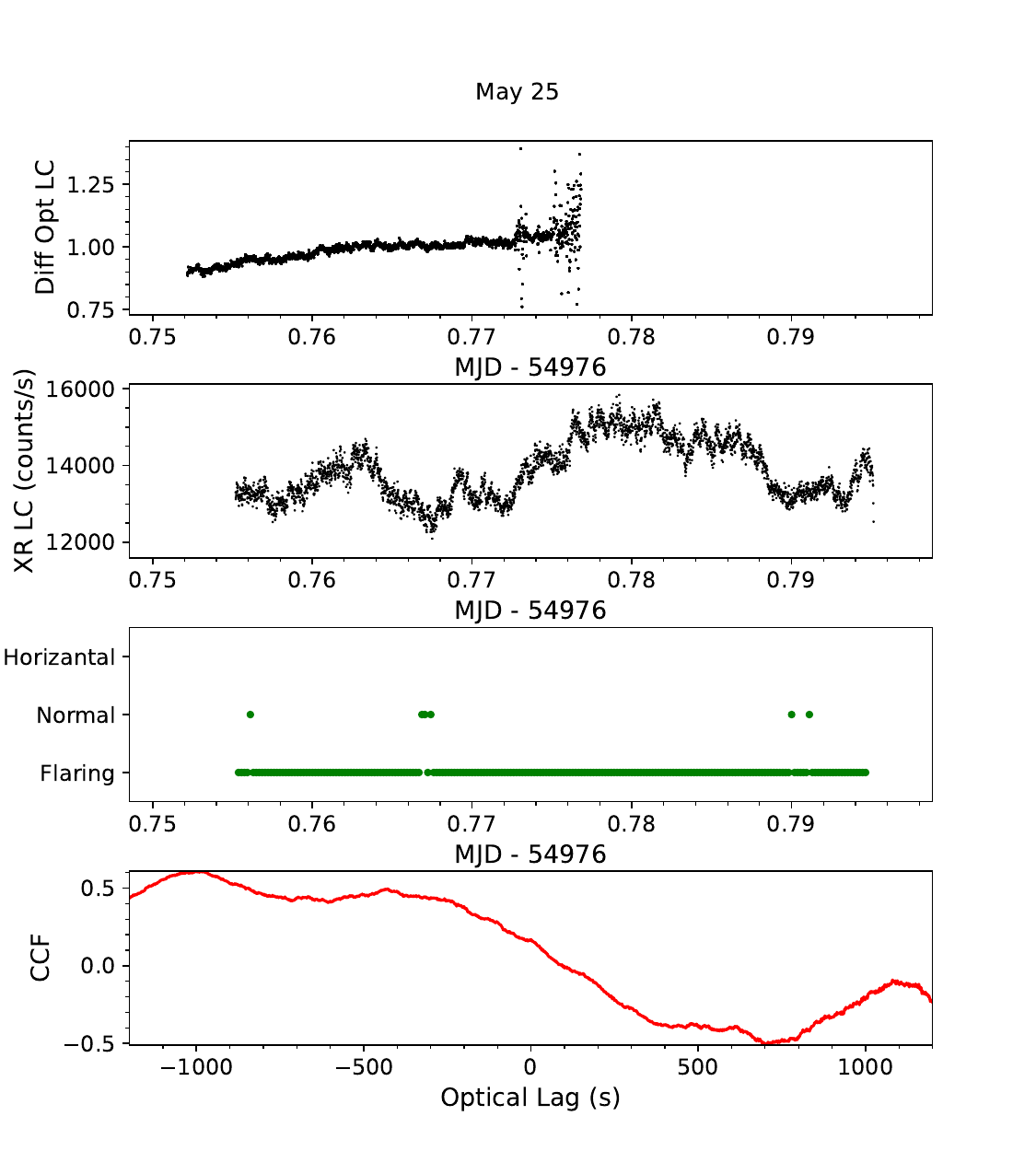}
        \caption[empty]{}
        \label{fig:may25a_full}
    \end{subfigure}
    \hfill
    \begin{subfigure}[b]{.475\textwidth}
        \setcounter{subfigure}{4}
        \includegraphics[width=\textwidth]{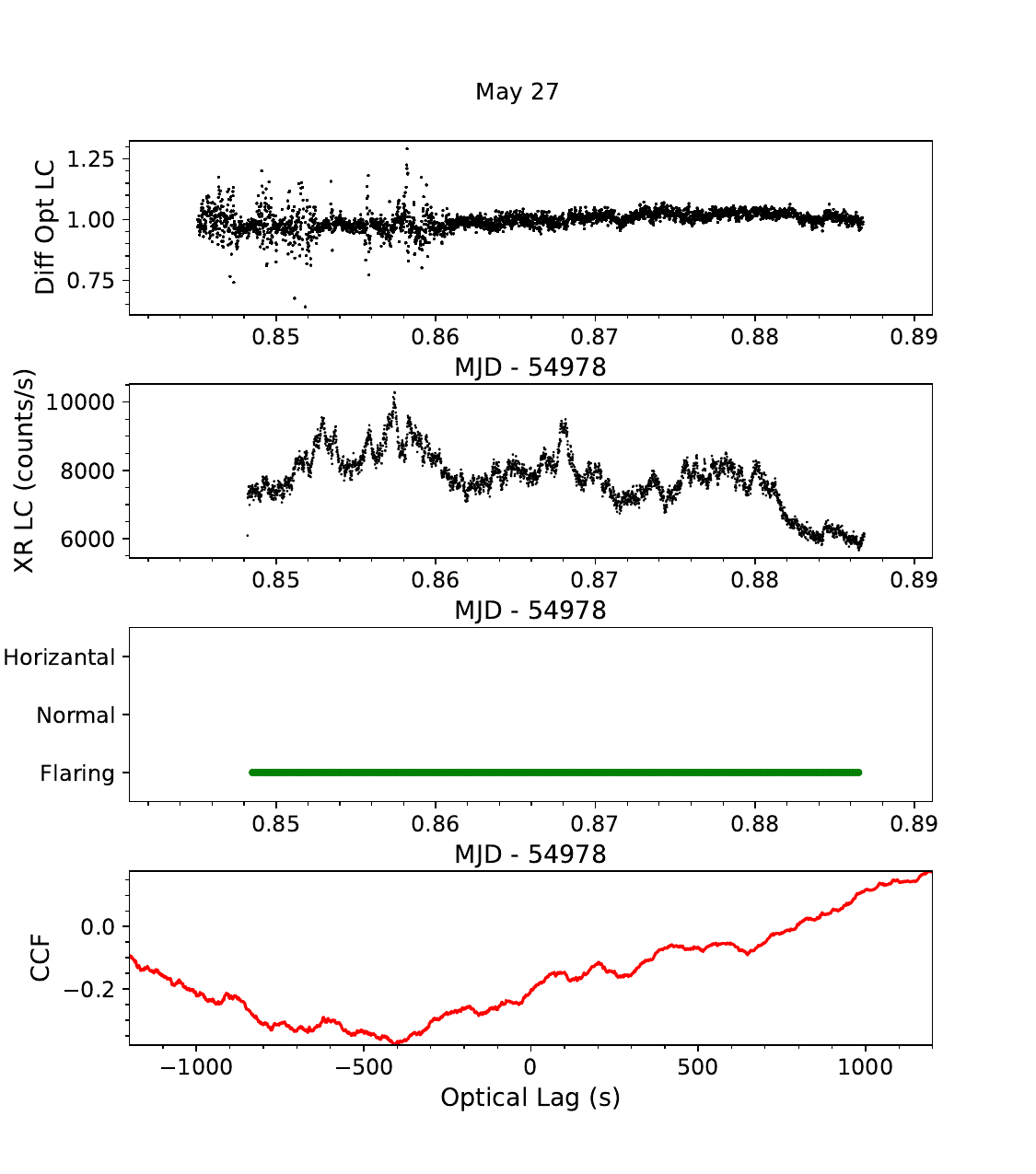}
        \caption[empty]{}
        \label{fig:may27b_full}
    \end{subfigure}
    \vspace*{5mm}
    \begin{subfigure}[b]{.475\textwidth}
        \setcounter{subfigure}{3}
        \includegraphics[width=\textwidth]{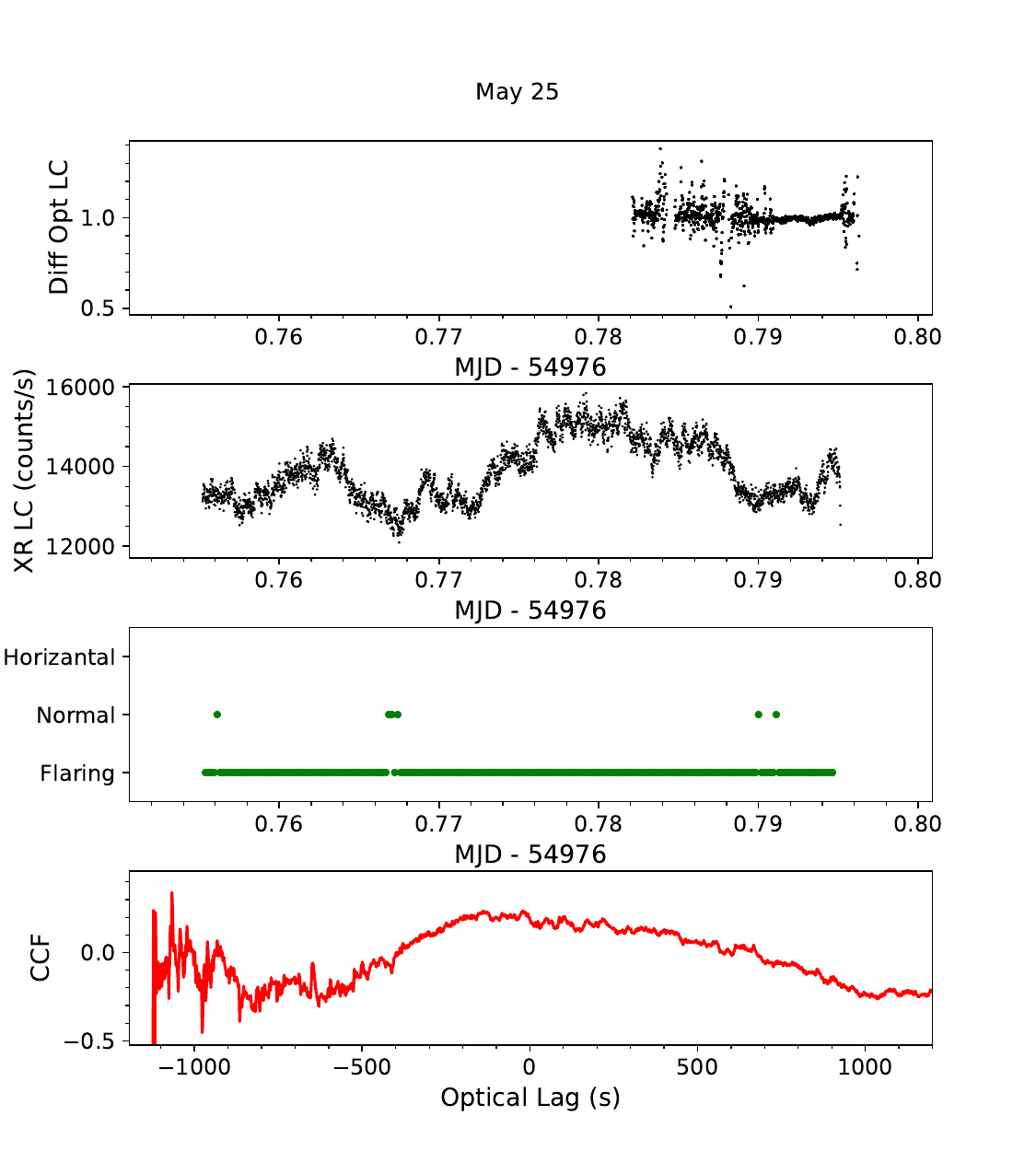}
        \caption[empty]{}
        \label{fig:may25b_full}
    \end{subfigure}
    \hfill
    \begin{subfigure}[b]{.475\textwidth}
        \setcounter{subfigure}{5}
        \includegraphics[width=\textwidth]{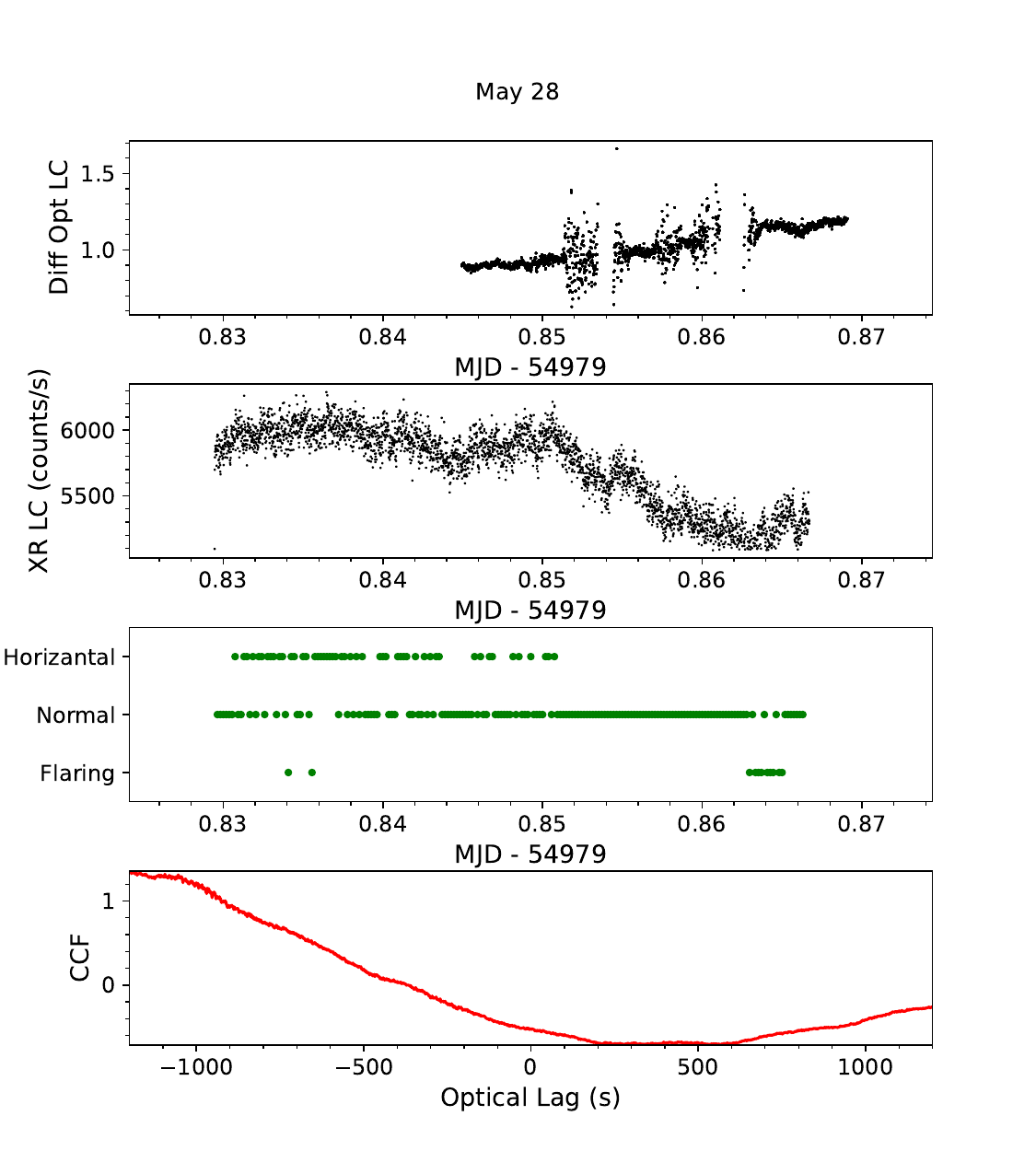}
        \caption[empty]{}
        \label{fig:may28a_full}
    \end{subfigure}
    \caption{(cont.) The first plot is the differential optical lightcurve (normalized by the median), the second is the X-ray intensity, the third is the location on the Z track, and the last shows the unfiltered CCF.}
\end{figure*}

\begin{figure*}
    \ContinuedFloat
    \begin{subfigure}[b]{.475\textwidth}
        \setcounter{subfigure}{6}
        \includegraphics[width=\textwidth]{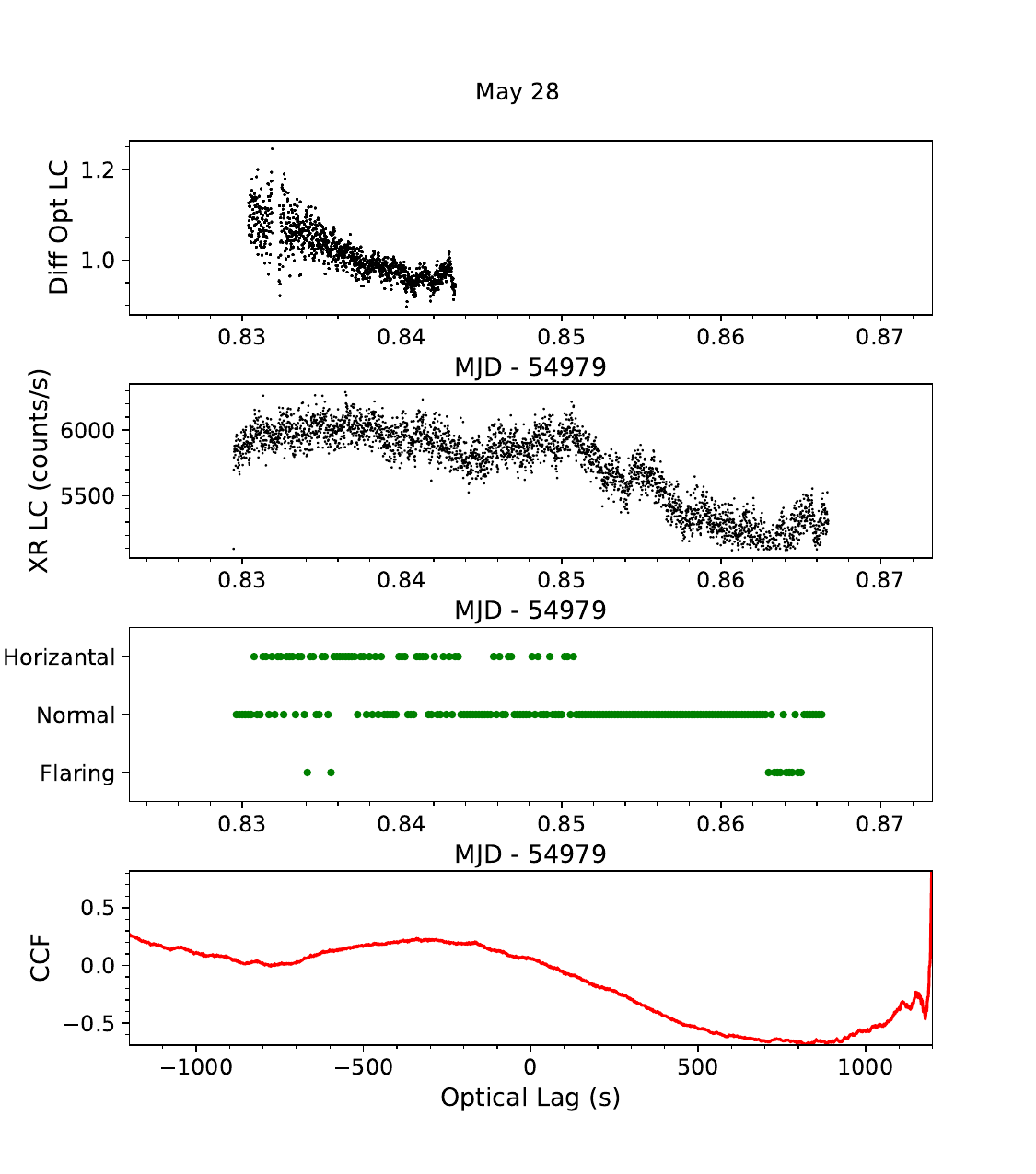}
        \caption[empty]{}
        \label{fig:may28b_full}
    \end{subfigure}
    \hfill
    \begin{subfigure}[b]{.475\textwidth}
        \setcounter{subfigure}{8}
        \includegraphics[width=\textwidth]{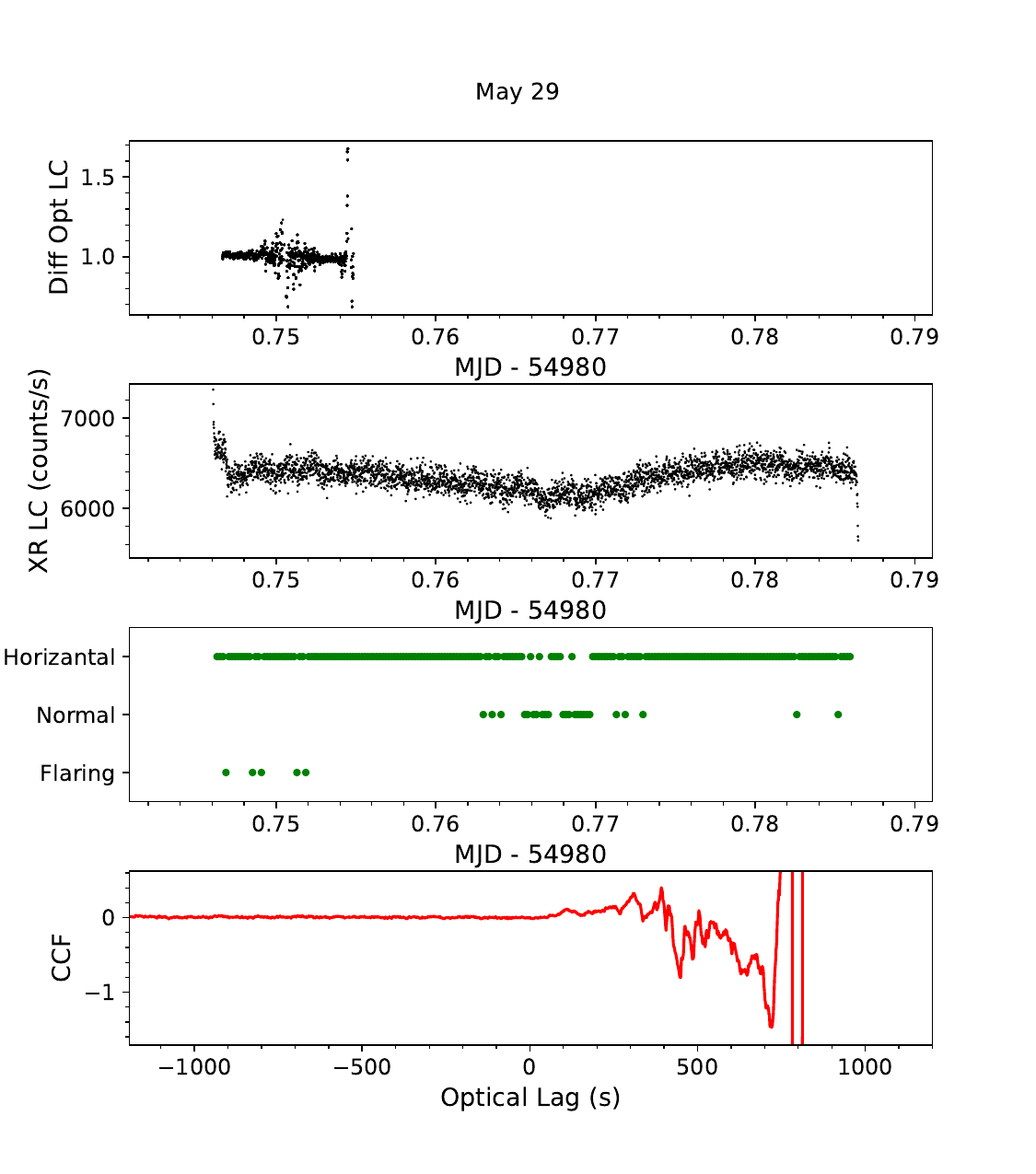}
        \caption[empty]{}
        \label{fig:may29b_full}
    \end{subfigure}
    \vspace*{5mm}
    \begin{subfigure}[b]{.475\textwidth}
        \setcounter{subfigure}{7}
        \includegraphics[width=\textwidth]{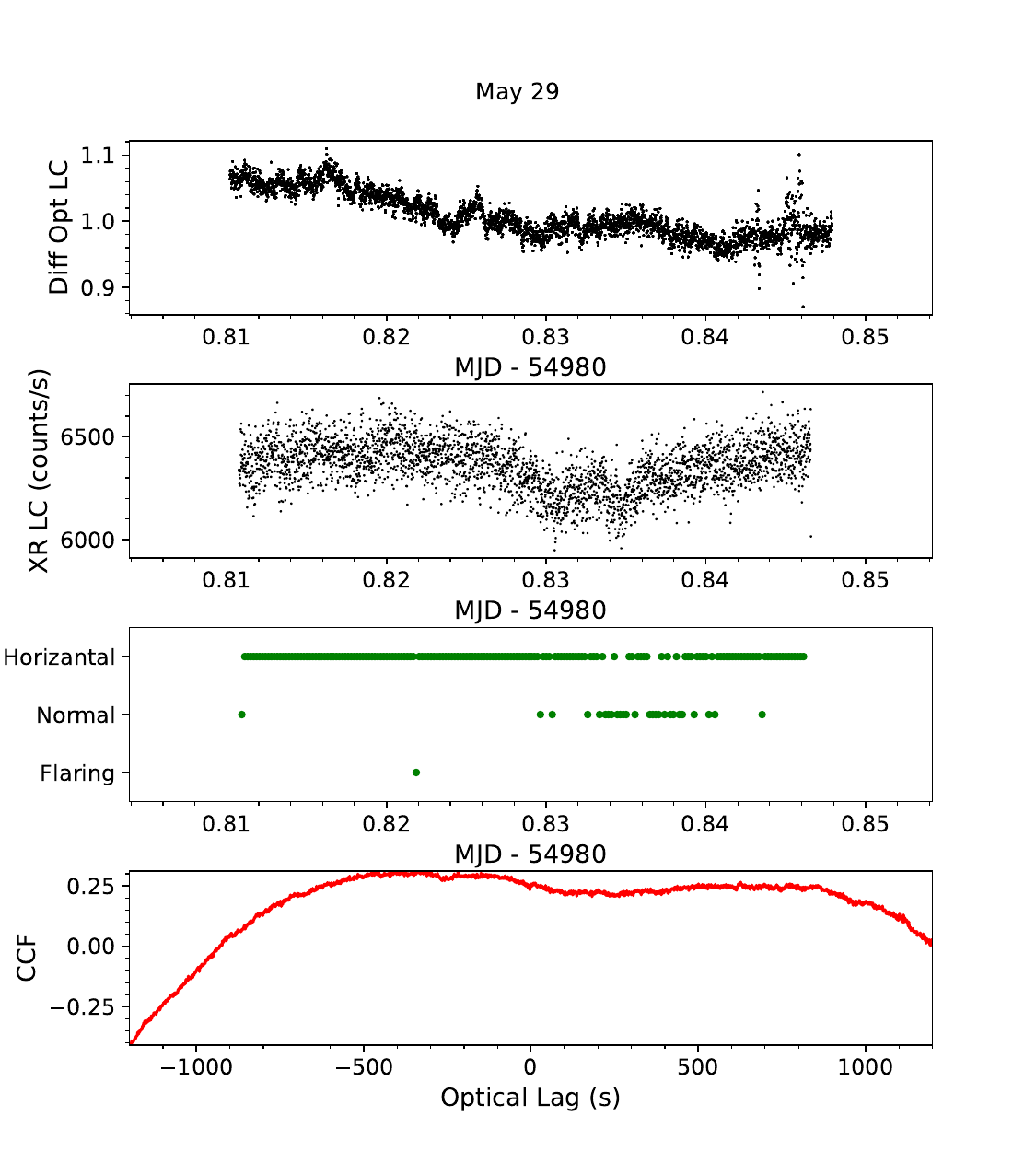}
        \caption[empty]{}
        \label{fig:may29a_full}
    \end{subfigure}
    \hfill
    \begin{subfigure}[b]{.475\textwidth}
        \setcounter{subfigure}{9}
        \includegraphics[width=\textwidth]{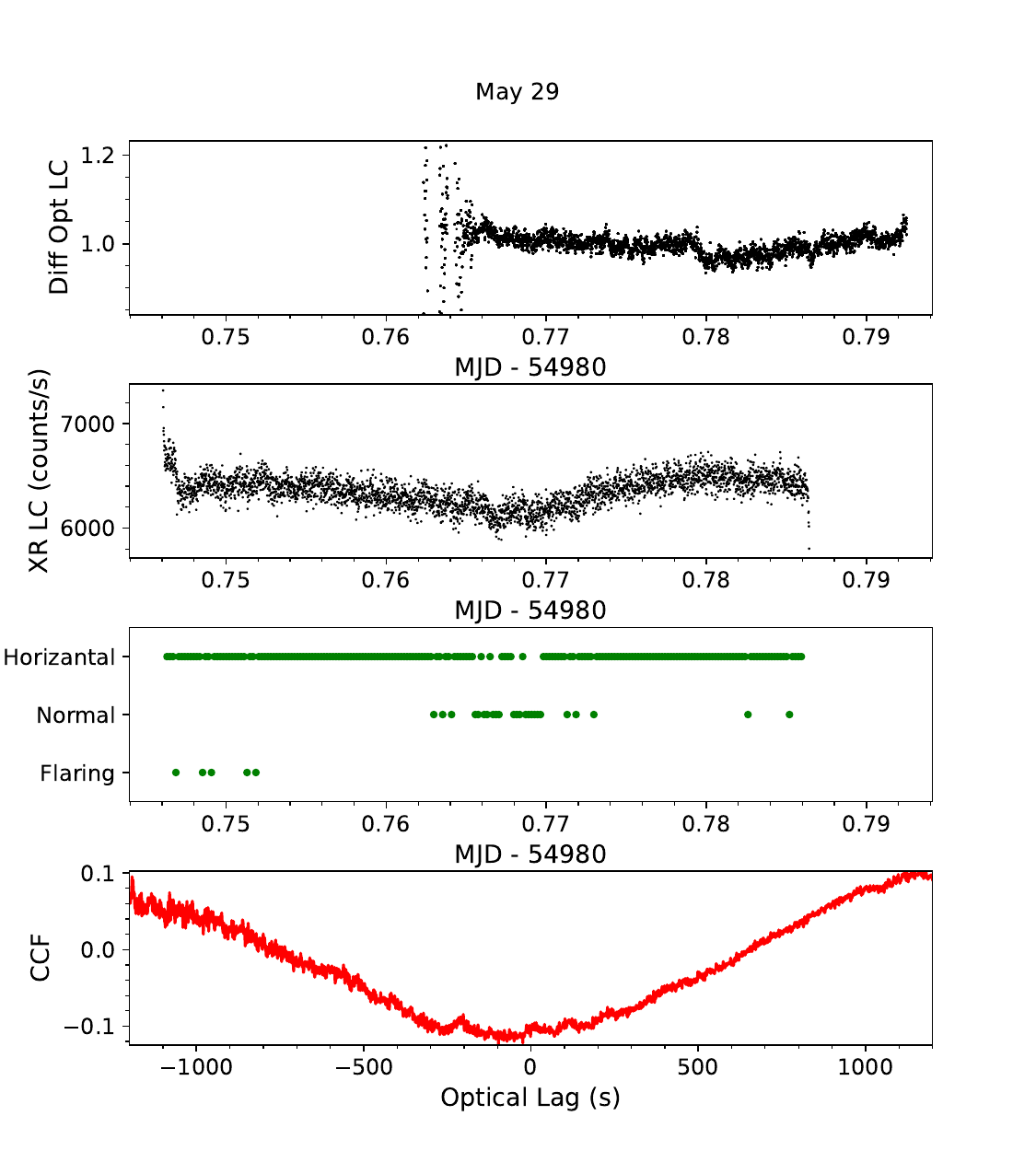}
        \caption[empty]{}
        \label{fig:may29c_full}
    \end{subfigure}
    \caption{(cont.) The first plot is the differential optical lightcurve (normalized by the median), the second is the X-ray intensity, the third is the location on the Z track, and the last shows the unfiltered CCF.}
\end{figure*}

\begin{figure}
    \ContinuedFloat
    \begin{subfigure}[b]{.475\textwidth}
        \setcounter{subfigure}{10}
        \includegraphics[width=\textwidth]{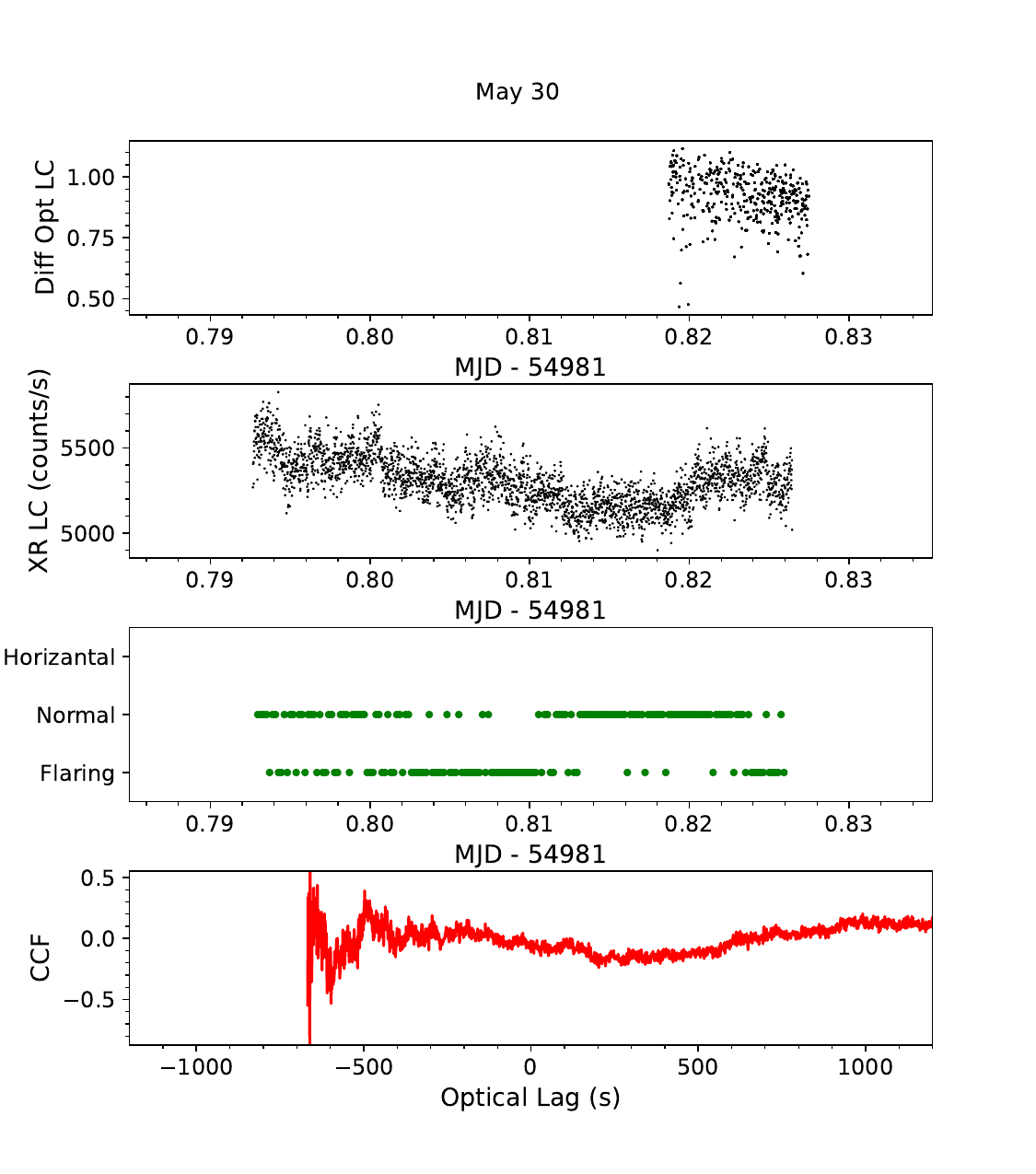}
        \caption[empty]{}
        \label{fig:may30a_full}
    \end{subfigure}
    \hfill
    \begin{subfigure}[b]{.475\textwidth}
        \setcounter{subfigure}{11}
        \includegraphics[width=\textwidth]{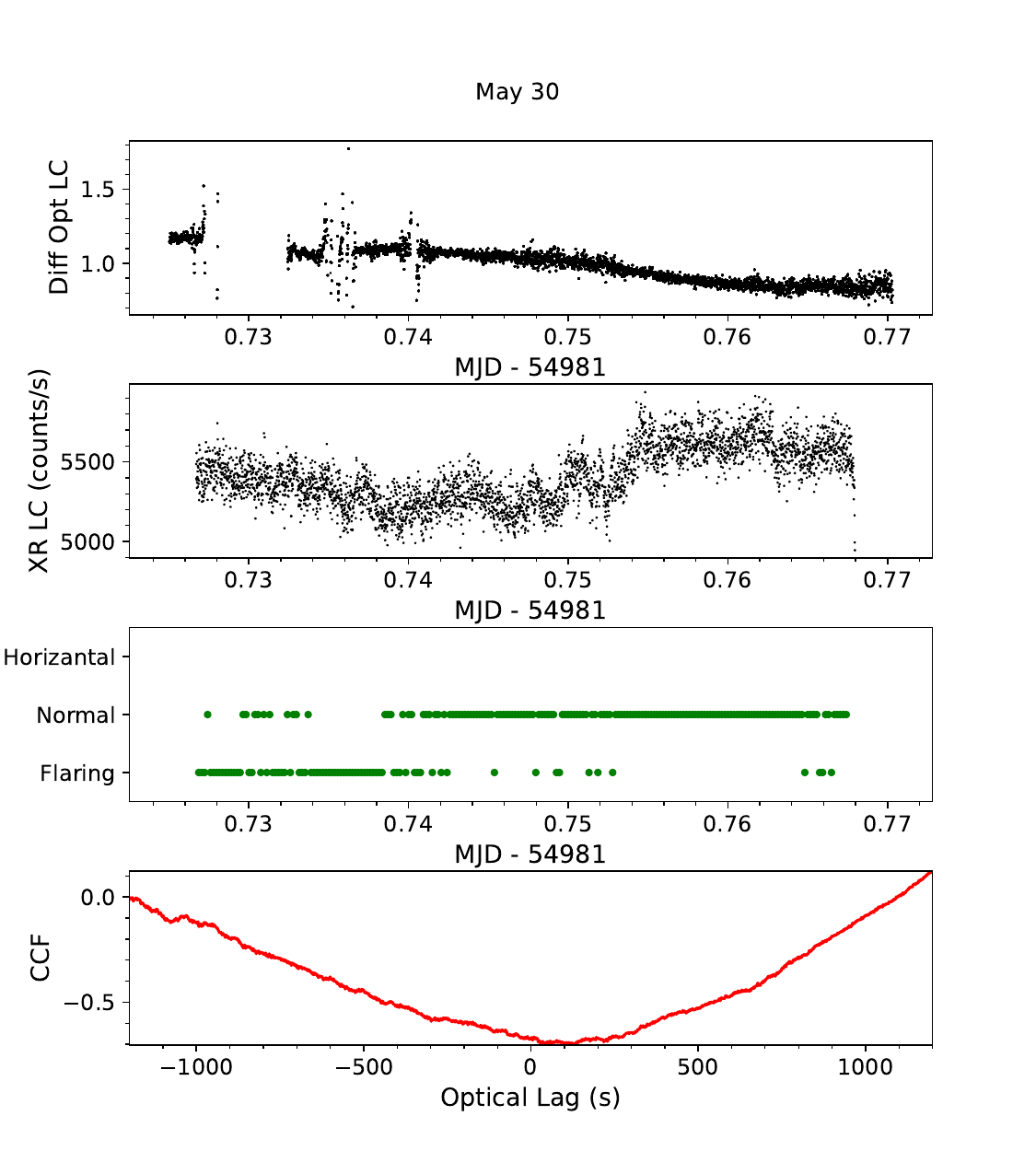}
        \caption[empty]{}
        \label{fig:may30b_full}
    \end{subfigure}
    \caption{(cont.) The first plot is the differential optical lightcurve (normalized by the median), the second is the X-ray intensity, the third is the location on the Z track, and the last shows the unfiltered CCF.}
\end{figure}

%%%%%%%%%%%%%%%%%%%%%%%%%%%%%%%%%%%%%%%%%%%%%%%%%%

% Don't change these lines
\bsp	% typesetting comment
\label{lastpage}
\end{document}